\documentclass[aps,preprint,superscriptaddress,groupedaddress,nofootinbib]{revtex4}

\usepackage{graphicx}
\usepackage[normalem]{ulem}
\usepackage{float}
\usepackage{amsmath}
\usepackage{amssymb}
\usepackage{booktabs} 
\usepackage{hhline}
\usepackage{multirow}
\usepackage[rightcaption]{sidecap}
\usepackage{hyperref}
\usepackage{xcolor}
\usepackage{wrapfig}

\graphicspath{ {./Figures/} }

\def\beq{\begin{equation}}
\def\eeq{\end{equation}}
\def\bea{\begin{eqnarray}}
\def\eea{\end{eqnarray}}

\def\figureautorefname~#1\null{Fig.\,#1\null}
\def\tableautorefname~#1\null{Tab.\,#1\null}

\def\equationautorefname~#1\null{Eq.\,(#1)\null}
\allowdisplaybreaks[4]

\begin{document}

\title{Primordial Black Holes from \\First-Order Phase Transition in the xSM}

\author{Dorival Gon\c{c}alves}
\email{dorival@okstate.edu}
\affiliation{Department of Physics, Oklahoma State University, Stillwater, OK, 74078, USA}
\author{Ajay Kaladharan}
\email{kaladharan.ajay@okstate.edu}
\affiliation{Department of Physics, Oklahoma State University, Stillwater, OK, 74078, USA}
\author{Yongcheng Wu}
\email{ycwu@njnu.edu.cn}
\affiliation{Department of Physics and Institute of Theoretical Physics, Nanjing Normal University, Nanjing, 210023, China}


\begin{abstract}
Supercooled first-order phase transition (FOPT) can lead to the formation of primordial black holes (PBHs). This scenario imposes stringent requirements on the profile of the effective potential. In this work, we use the singlet extended Standard Model (xSM) as a benchmark model to investigate this possibility at the electroweak scale. The PBHs formed during a supercooled FOPT have a narrow mass distribution around the mass of Earth. This distribution is closely tied to the temperature at which the PBHs form, corresponding to the FOPT at the electroweak scale. This scenario can be probed with microlensing experiments, space-based gravitational wave detectors, and collider experiments. Remarkably, the future space-based gravitational wave detector LISA will hold the potential to either confirm this scenario that leads to PBH formation across the observable Universe in the xSM or completely rule it out.
Interestingly, our findings suggest that PBHs within the xSM framework may align with observations of the six ultrashort timescale events reported by the OGLE microlensing experiment.
\end{abstract}



\maketitle
\flushbottom
\clearpage
\tableofcontents
\clearpage
\section{Introduction}
\label{sec:intro}

\begin{figure}[b!]
    \includegraphics[width=0.85\textwidth]{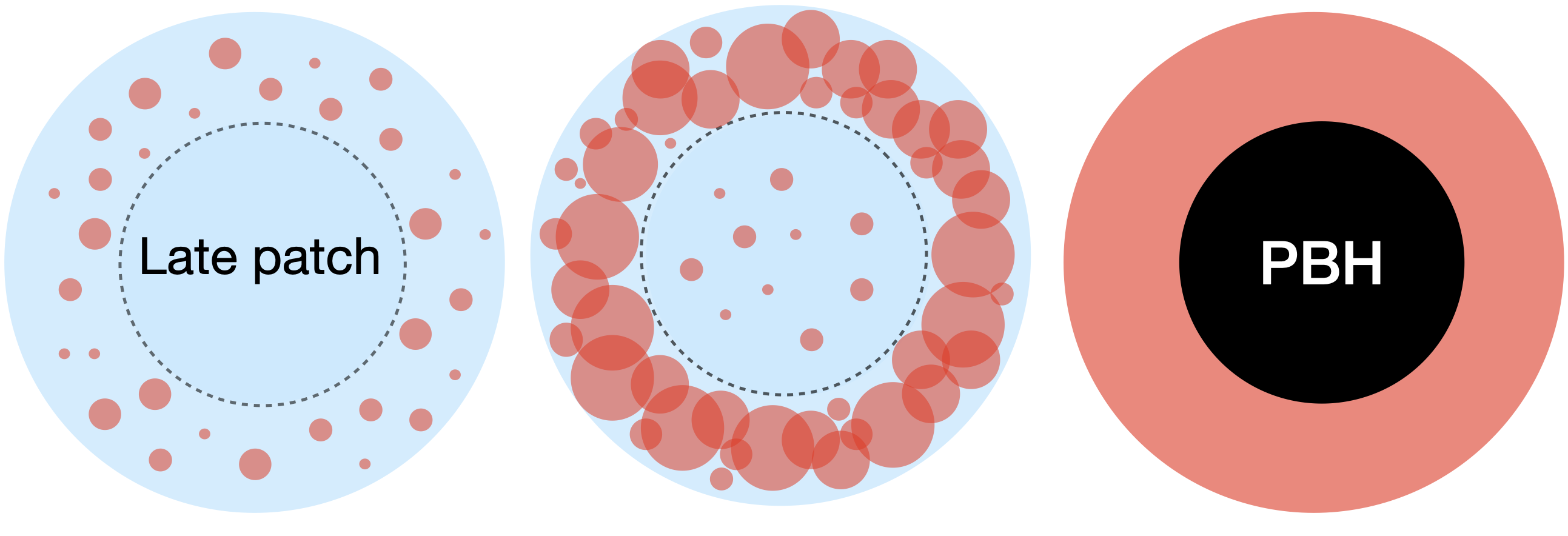}
    \caption{PBH formation from collapse of delayed false vacuum patches. Left panel: Bubble nucleation, occurring randomly, may start later in certain causal patches compared to the surrounding regions. Central panel: In supercooled conditions,  false vacuum regions in blue are vacuum-dominated, while true vacuum regions in red are energetically dominated by components that redshift akin to radiation. Thus, the background redshifts quickly, whereas the ``late patch" maintains a nearly constant energy density. Right panel: As a result of the delayed nucleation of bubbles in the late patch, energy density within it surpasses that of the background region.  When this overdensity crosses a critical threshold, it can collapse to form a PBH.}
    \label{fig:sketch}
\end{figure}

Primordial black holes (PBHs) can be formed from the collapse of large density inhomogeneities in the early Universe~\cite{Hawking:1971ei,Carr:1974nx, 1975ApJ...201....1C,Khlopov:1998nm,Dent:2024yje,Bertuzzo:2024fns}.
Cosmological first-order phase transition may provide ideal conditions for PBH formation through the compression of large amount of energy into a small volume through bubble collisions~\cite{PhysRevD.26.2681,PhysRevD.50.676,Lewicki:2019gmv},
trapped particles~\cite{Baker:2021nyl,Baker:2021sno,Huang:2022him,Gehrman:2023esa,Lewicki:2023mik,Gehrman:2023qjn}, or delayed vacuum transition~\cite{10.1143/PTP.68.1979,Liu:2021svg,Hashino:2021qoq, Hashino:2022tcs,Kawana:2022olo, Lewicki:2023ioy,Gouttenoire:2023naa,Baldes:2023rqv,Salvio:2023ynn,Gouttenoire:2023pxh,Gouttenoire:2023bqy,Conaci:2024tlc,Flores:2024lng, Lewicki:2024ghw,Kanemura:2024pae}. In this paper, we will focus on the latter scenario.

The probabilistic nature of bubble nucleation implies that there is the possibility that large regions may remain in the false vacuum state, where nucleation is delayed, and are surrounded by regions dominated by the true vacuum state.  While the vacuum energy density remains almost constant, the radiation energy density decreases as the Universe expands. Therefore, the total energy density increases in the region where false vacuum decay is delayed relative to the background region. When the vacuum energy density inside those regions decays into other components, overdensity may reach a threshold, and the whole mass inside the region could gravitationally collapse into PBHs. In this scenario, PBHs can serve as both dark matter candidates and probe for models featuring first-order phase transitions, making it phenomenologically appealing, as we will illustrate in the present study. In Fig.~\ref{fig:sketch}, we summarize the process for PBH formation from collapse of delayed false vacuum patches.

In this work, we will demonstrate, for the first time,  that this scenario of PBH formation can arise in the context of the relevant theoretical benchmark known as xSM. The xSM extends the Standard Model (SM) by introducing a new gauge singlet scalar. Our analysis reveals that this model can generate a wide range of dark matter fractions.  Furthermore, the PBH mass spectrum in this model is highly constrained around $10^{-5}$ solar masses, reflecting a direct correlation with the temperature at which the PBHs form.
Interestingly, the current OGLE experiment has detected six ultrashort timescale events that align with the characteristics expected of Earth-mass PBHs~\cite{Niikura:2019kqi,Scholtz:2019csj}. Thus, PBH formation via a supercooled electroweak phase transition emerges as a potential explanation.

We will show that the PBH phenomenology will lead to novel and relevant constraints on this model from microlensing experiments, complementing constraints from collider experiments such as the High-Luminosity LHC (HL-LHC)~\cite{Profumo:2007wc,Robens:2015gla, Huang:2017jws, Chen:2017qcz, Goncalves:2018qas,Alves:2018jsw,Alves:2018oct,Biekotter:2018jzu,Alves:2019igs,Profumo:2014opa,Barman:2020ulr,Alves:2020bpi,Goncalves:2021egx,Goncalves:2022wbp,Anisha:2022hgv,Palit:2023dvs,Goncalves:2023svb}, and future gravitational wave (GW) experiments such as the Laser Interferometer Space Antenna (LISA)~\cite{amaroseoane2017laser}. Therefore, we will illustrate using the xSM, that models featuring a first-order electroweak phase transition can be comprehensively tested through a multipronged approach, including PBH phenomenology.

The remainder of the paper is organized as follows. We briefly introduce the xSM model in~\autoref{sec:model}.
The electroweak phase transition in the xSM is discussed in~\autoref{sec:EWPT}. Moving to~\autoref{sec:PBH_PT}, we analyze the mechanism of PBH formation during a delayed phase transition.  The results in the xSM model will be presented in~\autoref{sec:Results}. We start with a general discussion about the requirements of PBH formation to the structure of the Higgs potential. This is followed by the presentation of constraints from microlensing, gravitational waves, and collider experiments. We conclude in~\autoref{sec:conclusion}. Details about the parameter space scan, theoretical and experimental constraints to the xSM, PBH calculation, and the calculation of GW signal are provided in the Appendices.

\section{The xSM model}
\label{sec:model}
In the xSM framework, the SM is extended by one real scalar field $S$, which is a singlet under the SM symmetry $\mathrm{SU}(3)_\mathrm{C} \otimes \mathrm{SU}(2)_\mathrm{L} \otimes \mathrm{U}(1)_\mathrm{Y}$~\cite{Robens:2015gla, Huang:2017jws, Alves:2018oct, Chen:2017qcz, Alves:2019igs, Alves:2018jsw, Profumo:2007wc, Profumo:2014opa}. The general gauge-invariant scalar potential is given by
\begin{align}
V(H,S)=&-\mu^2 H^\dagger H+\lambda (H^\dagger H)^2+\frac {a_1}{2}H^\dagger HS \nonumber\\
&+\frac {a_2}{2}H^\dagger H S^2+\frac {b_2}{2}S^2+\frac{b_3}{3}S^3+\frac{b_4}{4}S^4,
\label{eq:V0}
\end{align}
where $H$ is the SM Higgs doublet, and all the model parameters in~\autoref{eq:V0} are real. Expanding around the vacuum expectation values (VEVs), the Higgs doublet $H$ and real scalar $S$ can be written as
\begin{equation}
H = \begin{pmatrix}
G^+\\
\frac{v_{\mathrm{EW}} +h + i G^0}{\sqrt{2}}
\end{pmatrix}, \hspace{0.5cm} \text{and} \hspace{0.5cm}S=v_\mathrm{s}+s,
\end{equation}
where $G^{\pm}$ and $G^0$ are the Goldstone fields, $h$ and $s$ are the Higgs and singlet scalars,  $v_{\mathrm{EW}}= 246\,\mathrm{GeV}$, and $v_s$ is the singlet VEV at zero-temperature. The minimization conditions around the electroweak (EW) vacuum at zero temperature allow parameters $\mu$ and $b_2$ in \autoref{eq:V0} to be expressed in terms of the VEVs and other parameters as
\begin{align}
\mu^2&=\lambda v^2_{\mathrm{EW}}+\frac 12 v_{\mathrm{s}}(a_1+a_2v_{\mathrm{s}}),\\
b_2&=-\frac{1}{4v_{\mathrm{s}}}\left [ v_{\mathrm{EW}}^2(a_1+2a_2 v_{\mathrm{s}})+4v_{\mathrm{s}}^2(b_3+b_4 v_{\mathrm{s}}) \right ].
\label{eq:min}
\end{align}
Using \autoref{eq:min}, the mass matrix for $(h,s)$ can be written as
\begin{equation}
M^2_{h,s}=\begin{pmatrix}
2\lambda v^2_{\mathrm{EW}} & \frac {1}{2}a_1v_{\mathrm{EW}}+a_2v_{\mathrm{EW}}v_{\mathrm{s}} \\
\frac {1}{2}a_1v_{\mathrm{EW}}+a_2v_{\mathrm{EW}}v_{\mathrm{s}} &  v_{\mathrm{s}}(b_3+2b_4v_{\mathrm{s}})-\frac {v_{\mathrm{EW}}^2}{4v_{\mathrm{s}}}a_1
\end{pmatrix}.
\label{eq:massmatrix}
\end{equation}
The physical eigenstate $(h_1, h_2)$ can be expressed in terms of gauge eigenstate $(h,s)$ and mixing angle $\theta$ which diagonalizes the matrix in \autoref{eq:massmatrix}\footnote{Here $s_x\equiv\sin(x)$ and $c_x\equiv\cos(x)$.}
\begin{equation}
h_1=c_\theta h+s_\theta s, \quad h_2=-s_\theta h+c_\theta s,
\label{eq:mixing}
\end{equation}
where $h_1$ is identified as the $125\,\mathrm{GeV}$ SM Higgs boson while $h_2$ is a new scalar.

The model parameters $(\lambda, a_1,a_2)$ can be expressed in terms of physical parameters $(m_{h_1}\, m_{h_2},\theta)$ as
\begin{align}
\lambda&=\frac {m^2_{h_1}c_{\theta}^2+m^2_{h_2}s_{\theta}^2}{2v_{\mathrm{EW}}^2}\,,\\
a_1&=\frac {2v_{\mathrm{s}}}{v_{\mathrm{EW}}^2}[2v_{\mathrm{s}}^2(2b_4+\tilde{b}_3)-m^2_{h_1}-m^2_{h_2}+c_{2\theta}(m^2_{h_1}-m^2_{h_2})]\,,\\
a_2&=\frac{-1}{2v_{\mathrm{EW}}^2v_{\mathrm{s}}}\left [-2v_{\mathrm{s}}(m^2_{h_1}+m^2_{h_2}-4b_4v_{\mathrm{s}}^2 )+(m^2_{h_1}-m^2_{h_2})(2c_{2\theta}v_{\mathrm{s}}-v_{\mathrm{EW}}s_{2\theta}) +4\tilde{b}_3v_{\mathrm{s}}^3 \right ]\,,
\end{align}
where $\tilde{b}_3\equiv b_3/v_{\mathrm{s}}$. Therefore, the model can be fully specified using the following five free parameters:
\begin{equation}
v_\mathrm{s},\quad m_{h_2},\quad\theta,\quad b_3,\quad b_4.
\end{equation}
The details of the parametrization scan are provided in Appendix~\ref{app:paramscan}.

By virtue of \autoref{eq:mixing}, the gauge and Yukawa couplings of $h_1$ and $h_2$ are scaled by a factor of $c_\theta$ and $s_\theta$ respectively with respect to the SM Higgs couplings
\begin{equation}
g_{h_1 XX}=c_\theta g_{hXX}^{\mathrm{SM}}, \quad \mathrm{and} \quad g_{h_2 XX}=-s_\theta g_{hXX}^{\mathrm{SM}},
\label{eq:ghxx}
\end{equation}
where $XX$ represents $W^+W^-, \, ZZ$, and $f\bar{f}$. The triple-Higgs coupling $\lambda_{111}h_1 h_1 h_1$ and the coupling between singlet and Higgs boson $\lambda_{211}h_2 h_1 h_1$ after electroweak symmetry breaking are given by~\cite{Huang:2017jws}
\begin{align}
\lambda_{111}=&\frac 12 \left [ -s^3_\theta\frac {b_3}{3}+s^2_\theta\left ( -\frac {v_{\mathrm{EW}}}{v_s}c_\theta+\frac 12 \frac {v_{\mathrm{EW}}^2}{v_s^2}s_\theta \right )\frac {a_1}{2}+m^2_{h_1}\left ( \frac {s^3_\theta}{v_s}+\frac {c_\theta^3}{v_{\mathrm{EW}}} \right ) \right ],
\label{eq:L111} \\
\lambda_{211}=&\frac{1}{4}[(a_1+2a_2v_{\mathrm{s}})c_{\theta}^3+4v_{\mathrm{EW}}(a_2-3\lambda)c_\theta^2 s_\theta \nonumber\\
&-2(a_1+2a_2v_{\mathrm{s}}-2b_3-6b_4v_{\mathrm{s}})c_{\theta}s_{\theta}^2-2a_2v_{\mathrm{EW}}s_\theta^3] .
\end{align}

Theoretical and experimental constraints for the xSM model used in this analysis are detailed in Appendix~\ref{app:constraints}.

\section{Electroweak phase transition}
\label{sec:EWPT}
The dynamics of electroweak symmetry breaking in the early Universe is governed by the effective potential. In the xSM, the barrier is generated primarily by tree-level cubic terms~\cite{Chung:2012vg, Goncalves:2021egx}. As tree-level terms drive the electroweak phase transition (EWPT), it is possible to perform a gauge invariant high-temperature expansion equivalent to considering only the thermal mass corrections~\cite{Patel:2011th,Wainwright:2011qy, Metaxas:1995ab, Garny:2012cg, Chiang:2017nmu, Arunasalam:2021zrs, Hirvonen:2021zej, Lofgren:2021ogg}. The gauge invariant effective potential is given by\footnote{ The inclusion of Coleman-Weinberg correction and full finite temperature one loop correction, including daisy resummation, would induce gauge dependency to the effective potential~\cite{Chiang:2017nmu, Patel:2011th,Wainwright:2011qy}.  However, by the Nielsen identities~\cite{Chiang:2017nmu}, the leading-order finite temperature correction,  $\mathcal{O}(T^2)$, remains gauge-independent. In this model, the barrier is predominantly generated by tree-level terms.
Hence, we opt to use the gauge-independent high-temperature approximation, in line with previous studies~\cite{Profumo:2007wc,Li:2019tfd,Alves:2019igs,Alves:2020bpi}. For the uncertainties associated with the calculations for EWPT, we refer to a more dedicated study~\cite{Croon:2020cgk}.}\,\footnote{In the $Z_2$ symmetric singlet scalar models, the cubic terms are absent, and one-loop Coleman-Weinberg terms are typically required to generate the barrier, which is essential for achieving strongly first-order electroweak phase transition (see, for instance~\cite{Carena:2019une}). As pointed out in~\cite{Carena:2019une}, the inclusion of the Coleman-Weinberg term can result in significant vacuum upliftment. In some cases, this may cause the electroweak symmetry-breaking vacuum to no longer be the global minimum at zero temperature, rendering such parameter points unphysical.}
\footnote{Ref.~\cite{Ramsey-Musolf:2024ykk} presented the calculation of the gauge-invariant two-loop effective potential for the EFT framework. The effects of two-loop contributions on PBH phenomenology in the xSM framework will be investigated in future work.}
\begin{align}
V_{\mathrm{eff}}(h,s,T)=&\frac 12[-\mu^2+\Pi_hT^2]h^2+\frac {1}{2}[b_2+\Pi_sT^2]s^2+\frac {\lambda}{4} h^4 \nonumber\\
&+\frac {a_1}{4}h^2 s+\frac {a_2}{4} h^2 s^2+\frac {b_3}{3}s^3+\frac{b_4}{4}s^4,
\end{align}
where $\Pi_hT^2$ and $\Pi_sT^2$ are the thermal masses of the fields, with coefficients given by
\begin{align}
\Pi_h=&\left ( \frac {2m^2_W+m^2_Z+m^2_t}{4v_{\mathrm{EW}}^2} +\frac {\lambda}{2}+\frac{a_2}{24}\right ),\\
\Pi_s=&\left ( \frac{a_2}{6}+\frac{b_4}{4} \right ),
\end{align}
where $m_W,\,m_Z$, and $m_t$ are the masses of $W$-boson, $Z$-boson and top quark, respectively. The xSM model exhibits both one-step and two-step phase transitions. In this study, we examine the potential PBH formation for both of these cases.

In the SM, electroweak symmetry breaking occurs via a smooth crossover transition~\cite{Kajantie:1996mn}. However, the introduction of new physics could convert it to a first-order phase transition. In this scenario, the phase transition proceeds through thermal tunneling from false to true vacua. This process leads to the emergence and expansion of bubbles of the broken phase within the surrounding region of the symmetric phase, effectively converting the false vacuum into true vacuum. The tunneling probability is provided by~\cite{Linde:1980tt,Coleman:1977py}
\begin{align}
    \Gamma (T)\approx T^4\left (\frac {S_3}{2\pi T}  \right )^{3/2}e^{-\frac {S_3}{T}}\,,
    \label{eq:Gamma}
\end{align}
where $S_3$ denotes the three-dimensional Euclidean action corresponding to the formation of the critical bubble
\begin{align}
    S_3=4\pi\int_{0}^{\infty}{dr r^2\left [ \frac 12\left ( \frac {d\phi(r)}{dr} \right )^2+V(\phi,T) \right ]}\,.
\end{align}
Here, the critical bubble profile $\phi$ is obtained by solving the following differential equation
\begin{align}
    \frac {d^2\phi}{dr^2}+\frac 2r \frac{d\phi}{dr}=\frac {dV(\phi,T)}{d\phi}\,, \quad   \text{with} \quad \lim_{r\rightarrow \infty}\phi(r)=0
    \quad \text{and} \quad \lim_{r\rightarrow 0}\frac {d\phi(r)}{dr}=0.
    \label{eq:tunneling}
\end{align}

We employ the publicly available code {\tt CosmoTransitions}~\cite{Wainwright:2011kj} to solve the differential equation~\autoref{eq:tunneling} and subsequently evaluate the Euclidean action $S_3$. The first-order phase transition is considered nearly complete around the nucleation temperature $T_n$,
at which one bubble nucleates per unit horizon volume~\cite{Moreno:1998bq}
\begin{align}
    \int_{T_n}^{\infty}\frac {dT}{T}\frac {\Gamma (T)}{H(T)^4}=1\,.
\label{eq:Tncond1}
\end{align}

Strong first-order phase transition can generate stochastic gravitational wave signals.
These GW signals emerge from three sources: vacuum bubble collision, fluid motion in the plasma resembling sound waves, and turbulent motion within the plasma. In this work, we are interested in the parameter space that leads to PBH formation, coinciding with a supercooling phase transition. In this scenario, it is more robust to consider the percolation temperature $T_p$,  when $29\%$ of space is covered by true vacuum bubbles, as the transition temperature for evaluating the gravitational signal~\cite{Athron:2022mmm,Athron:2023xlk,Athron:2023mer,Athron:2023rfq}.

One important parameter which influences the GW signal is $\alpha$, which denotes the ratio of vacuum energy density $\Lambda_{\mathrm{vac}}$ released as latent heat during phase transition to the radiation energy density $\rho_{\rm rad}$.{\footnote{For a more comprehensive analysis, the parameter $\beta$, commonly employed in modelling GW signals in the literature, is replaced by length scale $R_{\star}= (8\pi)^{\frac{1}{3}}\frac{v_w}{\beta}$, which represents the mean separation of  bubbles. The details of the evaluation of $R_{\star}$ are provided in Appendix~\ref{app:GW}.}} The vacuum energy density, radiation energy density, and $\alpha$ are given by
\begin{align}
\label{equ:energy_densities}
 \Lambda _{\mathrm{vac}} = \Delta\left(- V_{\rm eff} + T\frac{\partial V_{\rm eff}}{\partial T}\right), \quad \rho_{\rm rad} = \frac{\pi^2}{30}g_\star T^4,\,\quad \text{and} \quad \alpha=\left.\frac {\Lambda _{\mathrm{vac}}}{\rho_{\mathrm{rad}}}  \right |_{T=T_p},
\end{align}
where $g_\star=106.75$  denotes the number of relativistic degrees of freedom in the plasma and $\Delta$ denotes the difference between the true and false vacua. The calculation details for the GW energy density $\Omega_{\mathrm{GW}}h^2$  are provided in Appendix~\ref{app:GW}.

\section{Primordial black holes from first-order phase transition}
\label{sec:PBH_PT}

PBHs can be formed from the collapse of large density inhomogeneities in the early Universe~~\cite{Hawking:1971ei,Carr:1974nx, 1975ApJ...201....1C,Khlopov:1998nm}. In this study, we focus on PBH formation due to a delayed vacuum transition~\cite{10.1143/PTP.68.1979,Liu:2021svg,Hashino:2021qoq, Hashino:2022tcs,Kawana:2022olo, Lewicki:2023ioy,Gouttenoire:2023naa,Baldes:2023rqv,Salvio:2023ynn,Conaci:2024tlc,Flores:2024lng, Lewicki:2024ghw,Kanemura:2024pae}. The probabilistic nature of bubble nucleation implies that some large regions may remain longer in a false vacuum state, surrounded by areas that have transitioned to the true vacuum state. While the vacuum energy density remains approximately constant, the radiation energy density decreases as $a(t)^{-4}$ when the Universe expands. Hence, the total energy density in regions with delayed vacuum decay increases compared to surrounding regions where the decay occurs earlier. The resulting overdensity can reach a critical threshold, causing the entire mass of the region to gravitationally collapse into PBHs, as illustrated in~\autoref{fig:sketch}.

In general, the expected volume of true vacuum bubbles per comoving volume is given by~\cite{PhysRevD.46.2384}
\begin{equation}
I(t)=\frac {4\pi}{3}\int_{t_0}^{t}dt^\prime\Gamma(t^\prime)a^3(t^\prime)r^3(t,t^\prime),
\label{equ:I_t}
\end{equation}
where $t_0$ is the time when the first bubble is nucleated, and  $r(t,t^\prime)\equiv v_w \int_{t^\prime}^ta(t)^{-1}dt$ is the comoving radius of a bubble nucleated at time $t^\prime$ and observed at $t$.
We assume the velocity of the bubble wall close to the speed of light $v_w\approx1$.\footnote{The Chapman-Jouguet velocity~\cite{PhysRevD.25.2074,Giese:2020rtr} in~\autoref{eq:vw} sets the lower limit for the bubble wall velocity. For $\alpha = 1$ (10), the Chapman-Jouguet velocity is $0.934$ ($0.991$). In our investigation, we are more interested in the $\alpha\gg1$ regime, where the Chapman-Jouguet velocity approaches unity.}
$I(t)$ overestimates the fraction of true vacuum bubbles, as it double counts the overlap region and also includes fictitious nucleation in the true vacuum~\cite{PhysRevD.46.2384, PhysRevLett.44.631}.
The probability that a given point in the comoving volume remains in the false vacuum is given by~\cite{Ellis:2018mja,PhysRevD.46.2384, Liu:2021svg}
\begin{equation}
F(t) = \begin{cases}
1,         \, & \text{if $t< t_0$},\\
e^{-I(t)}, \, & \text{if $t\ge t_0$},
\end{cases}
\label{equ:false_fraction}
\end{equation}
where the exponential compensates the double counting mentioned above. Before the nucleation of the first bubble at $t_0$, the entire space is covered by a false vacuum; therefore, $F(t)=1$ for $t<t_0$ and $F(t)$ 
decreases as the true vacuum bubble nucleates and expands. With this probability, the percolation time $t_p$ (and equivalently the corresponding percolation temperature $T_p$) is defined as the point when the false vacuum fraction drops below the threshold $F(t_p)\simeq 0.71$.

The evolution of the Hubble scale factor is determined by the Friedmann equation~\cite{Friedmann:1924bb}
\begin{equation}
H^2=\left ( \frac 1 a\frac{\mathrm{d} a}{\mathrm{d} t} \right )^2=\frac {1}{3M_{\mathrm{Pl}}^2}(\rho_V+\rho_r+\rho_w),
\label{eq:Fried1}
\end{equation}
where $\rho_V,\, \rho_r,$ and $\rho_w$ are the energy densities for the false vacuum, background radiation, and bubble wall, respectively, and $M_{\mathrm{Pl}}=2.435 \times 10^{18}$~GeV~\cite{Ellis:2018mja}. The vacuum energy density $\rho_V$ is given by
\begin{equation}
\rho_V=F(t)\Lambda_{\mathrm{vac}}(t),
\end{equation}
where $\Lambda_{\mathrm{vac}}$ is the energy density difference between false and true vacuum at time $t$, as defined in~\autoref{equ:energy_densities}.
Given that the bubble wall velocity is assumed to be close to the speed of light, the bubble wall can be effectively treated as radiation~\cite{Gouttenoire:2023naa}. Therefore, the total radiation energy density is the sum of the background radiation and the bubble wall contributions
\begin{equation}
\rho_R=\rho_r+\rho_w,
\end{equation}
and the evolution of radiation energy density assuming energy-momentum conservation is described by
\begin{equation}
\frac {d\rho_R}{dt}+4H{\rho_R}=-\frac {d\rho_V}{dt}.
\label{eq:Fried2}
\end{equation}
In particular, \autoref{eq:Fried2} shows that as $F(t)$ decreases from 1 to 0, the vacuum energy is converted to radiation.

We consider a region where nucleation of the first bubble is delayed to time $t_i$, referred to as {\it late patch}.
The average spatial fraction of the false vacuum in this region is obtained from~\autoref{equ:I_t} and~\autoref{equ:false_fraction} with the delayed nucleation time
\begin{equation}
F_{\mathrm{in}}(t)=\begin{cases}
1, \, &t<t_i\\
\mathrm{exp}\left [ -\frac {4\pi}{3}\int_{t_i}^{t}dt^\prime\Gamma(t^\prime)a_{\mathrm{in}}^3(t^\prime)\left (\int _{t^\prime}^t\frac { d\tilde{t}}{a_{\mathrm{in}}(\tilde{t})} \right )^3 \right ], \, &t\ge t_i
\end{cases}
\label{eq:Fin}
\end{equation}
where $a_{\mathrm{in}}$ is the scale factor of the {\it late patch}.  Similarly, the fraction of false vacuum for the background region $F_{\mathrm{out}}$ can be evaluated with the nucleation time $t_0=t_c$. Due to the delayed nucleation of bubbles in the late patch, the energy density in the {\it late patch} is higher than in the background region. This energy contrast can be quantified as~\cite{Kawana:2022olo, Liu:2021svg}
\begin{equation}
\delta=\frac {\rho^{\mathrm{in}}-\rho^{\mathrm{out}}}{\rho^{\mathrm{out}}}.
\end{equation}
If the energy contrast exceeds the critical threshold, $\delta_c=0.45$~\cite{Harada:2013epa, Musco:2012au}, the {\it late patch} gravitationally collapses into PBH.
For a given $t_i$, we find the $t_{\mathrm{PBH}}$ time at which $\delta$ reaches the threshold value $\delta_c=0.45$ and PBH is formed.

The probability of the {\it late patch} remaining in the false vacuum until $t_i$ is given by~\cite{Kawana:2022olo, Liu:2021svg,Baldes:2023rqv}
\begin{equation}
P(t_i)=\mathrm{exp}\left [ -\int_{t_c}^{t_i}dt^\prime \Gamma(t^\prime) a_{\mathrm{in}}(t^\prime)^3V_{\mathrm{coll}} \right ],
\label{equ:P_pbh}
\end{equation}
where the volume factor $V_{\mathrm{coll}}$ is defined by~\cite{Kawana:2022olo}
\begin{equation}
V_{\mathrm{coll}}=\frac {4\pi}{3}\left [ \frac {1}{a_{\mathrm{in}}(t_{\mathrm{PBH}})H_{\mathrm{in}}(t_{\mathrm{PBH}})}+\int_{t^\prime}^{t_{\mathrm{PBH}}}\frac {d\tilde t}{a_{\mathrm{out}}(\tilde t)} \right ]^3.
\label{eq:Vcoll}
\end{equation}
 The second term in~\autoref{eq:Vcoll} represents the radius of the bubble nucleated at $t^\prime$ by the time $t_{\mathrm{PBH}}$. Including this term in~\autoref{eq:Vcoll} ensures that no large bubble nucleated in the surrounding background patch enters the {\it late patch} before its collapse at $t_{\mathrm{PBH}}$~\cite{Gouttenoire:2023naa}.

\begin{figure}[!tb]
\centering
\includegraphics[width=0.55\textwidth]{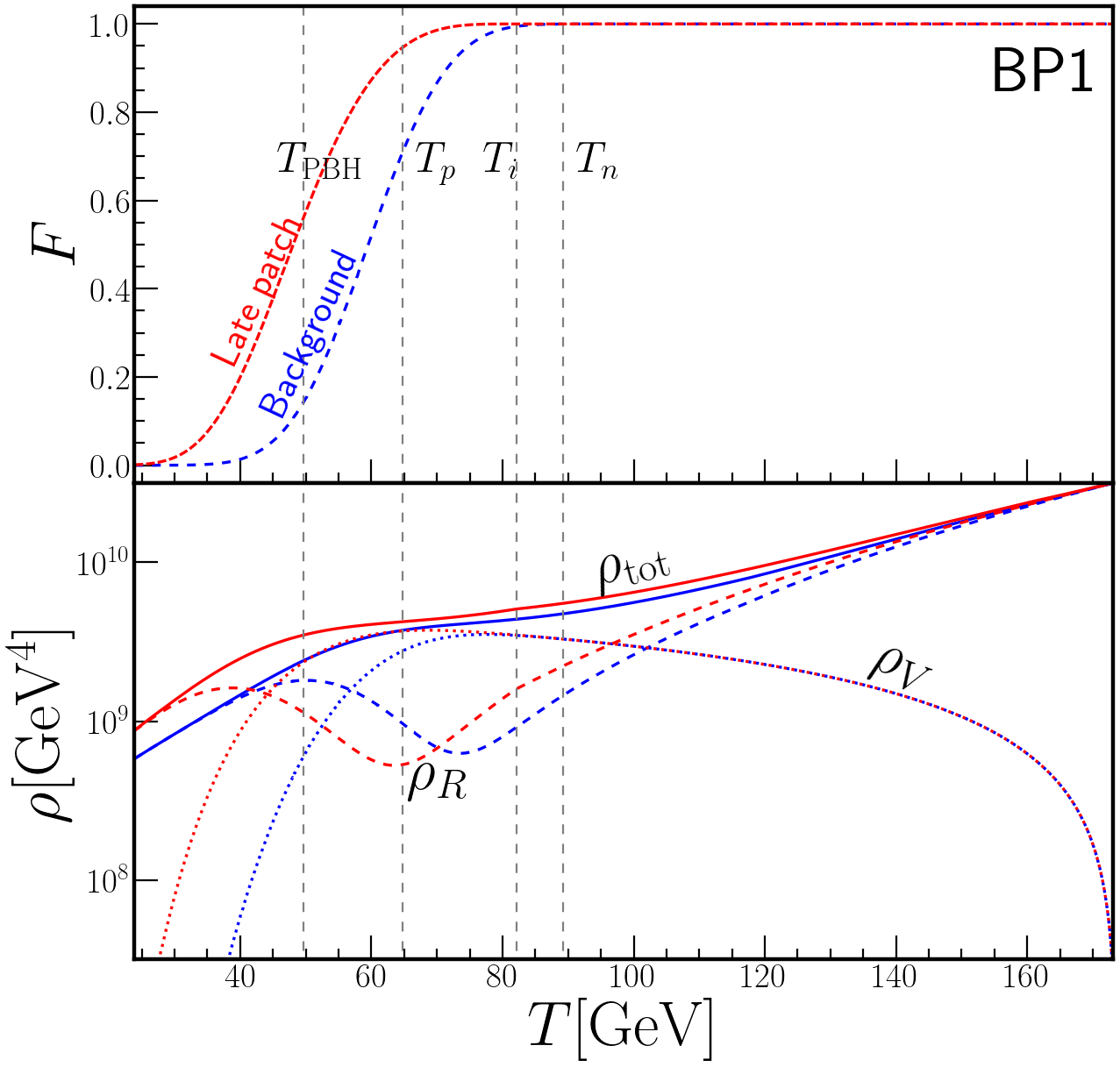}
\caption{In the top panel, we present the fraction of false vacuum inside and outside the late Hubble volume as a function of temperature for BP1, as defined in \autoref{tab:benchmark_points}.
 In the bottom panel, we show the evolution of energy density with temperature for both the background region and the late patch. In this panel, the dashed lines represent the radiation component, while the dotted lines indicate the vacuum energy density component.}
\label{BP1_Frac}
\end{figure}

The mass of PBH can be roughly approximated as the Hubble horizon mass at the temperature when the PBHs are produced~\cite{Liu:2021svg, Musco:2020jjb,Hashino:2021qoq, Hashino:2022tcs}
\begin{align}
M_{\mathrm{PBH}}\approx \frac {4\pi}{3} H_\mathrm{in}^{-3}(t_{\mathrm{PBH}})\rho_c={4\pi}{M_{\mathrm{Pl}}^2} H_\mathrm{in}^{-1}(T_{\mathrm{PBH}}),
\label{equ:MPBH}
\end{align}
where $\rho_c$ is the critical density. If we regard PBH as a component of dark matter (DM)~\cite{Green:2020jor}, the fraction of the PBH from the first-order phase transition in dark matter density is given by~\cite{Hashino:2021qoq, Hashino:2022tcs}
\begin{equation}
f_{\mathrm{PBH}}\equiv \frac{\rho_{\mathrm{PBH}}}{\rho_{\mathrm{DM}}}=\left ( \frac {H(t_{\mathrm{PBH}})}{H(t_0)} \right )^2\left ( \frac {a(t_\mathrm{PBH})}{a(t_0)} \right )^3P(t_i)\frac {1}{\Omega_{\mathrm{DM}}},
\label{eq:fpbh}
\end{equation}
where $t_0$ is the present time. For the electroweak phase transition $f_{\mathrm{PBH}}$ can be approximated as
\begin{equation}
f^{\mathrm{EW}}_{\mathrm{PBH}}\sim1.52 \times 10^{11}\left ( \frac {0.245}{\Omega_{\mathrm{DM}}} \right )\left ( \frac {T_{\mathrm{PBH}}}{100~\mathrm{GeV}} \right )P(T_i),
\end{equation}
where $\Omega_{\mathrm{DM}}=0.245$ is the present dark matter density normalized by total energy density.\footnote{PBHs with masses below $10^{-19}~M_{\odot}$ are expected to have evaporated by the present time due to Hawking radiation~\cite{Carr:2020gox}. However, PBHs formed during the electroweak transition typically have masses around $10^{-6}~M_{\odot}$. Estimates suggest that the impact of Hawking radiation on the mass of these PBHs is negligible. As a result, the overall mass of PBHs, and consequently their contribution to the dark matter density, remains largely unaffected.}$^{,}$\footnote{PBHs can accrete primordial gas in the early Universe, converting a fraction of this mass into radiation. This additional energy injection into the plasma can affect its thermal and ionization histories, leading to changes in the cosmic microwave background (CMB) frequency spectruum~\cite{10.1093/mnras/194.3.639}. These effects are subleading for PBHs formed during EWPT as considered in the present study, which typically have masses around $10^{-6}~M_\odot$, being relevant for PBHs with masses $\gtrsim 10^2~M_\odot$~\cite{Ali-Haimoud:2016mbv}.}

In~\autoref{BP1_Frac}, we illustrate the phenomenology described above using the xSM benchmark point 1 (BP1) listed in~\autoref{tab:benchmark_points}. The top panel compares the evolution of the false fraction vacuum with temperature in the {\it background region} $F_{\mathrm{out}}$, where nucleation begins at $T_n$, to the evolution in the {\it late patch} $F_{\mathrm{in}}$, where it is delayed to $T_i$.
The bottom panel depicts the evolution of the energy density for both the background and late patch regions. Both the background and late patch vacuum energies are converted into radiation, however, the background transitions earlier
making it to redshift quickly, whereas the late patch retains vacuum-dominated regions for a longer duration. This behavior results in an increase in $\delta$ as the vacuum decays in the late patch, ultimately reaching the PBH formation threshold at $\delta_c=0.45$, which defines $T_{\rm PBH}$.
\begin{figure}[tb!]
    \includegraphics[height=0.4\textwidth]{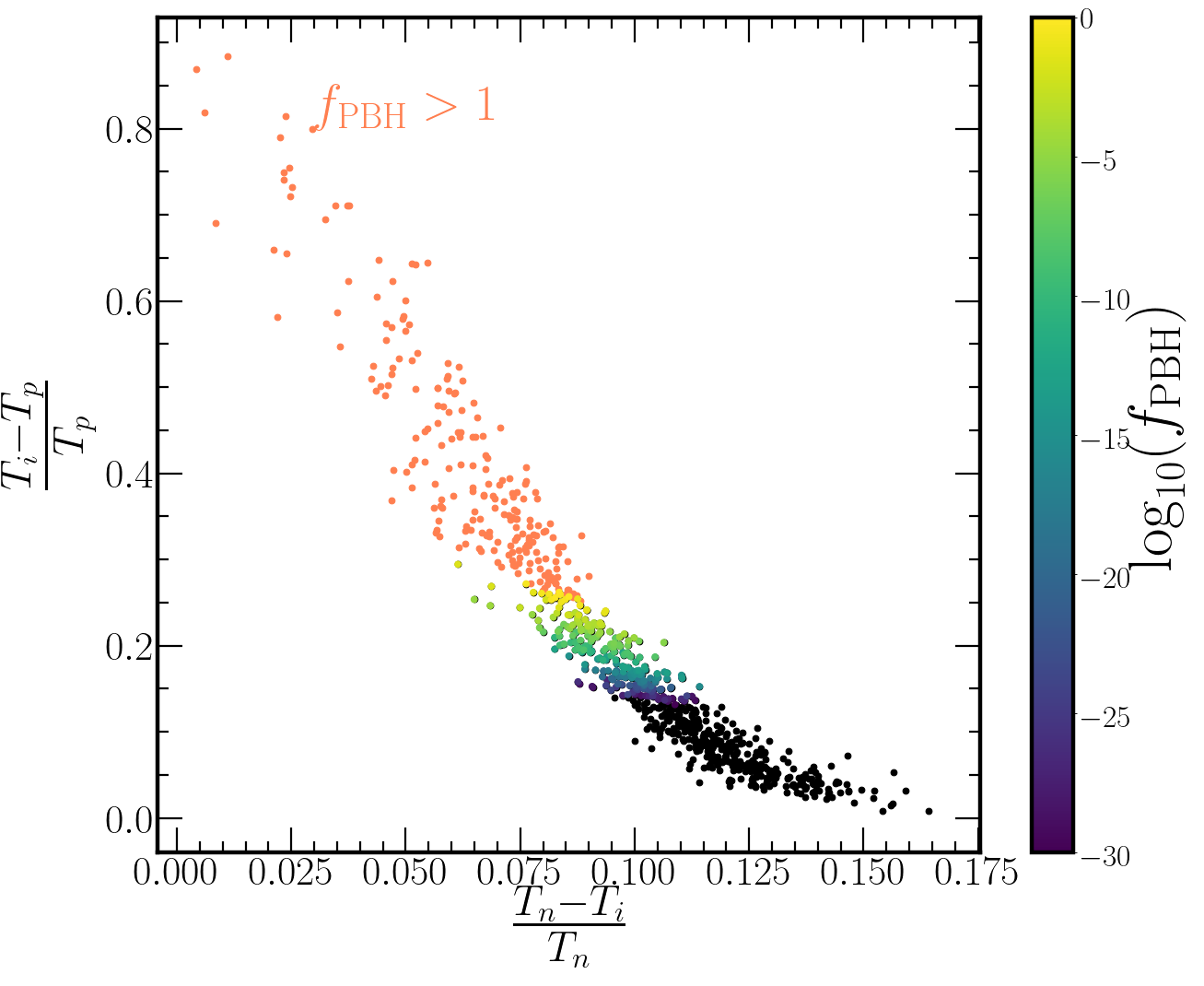}
    \includegraphics[height=0.4\textwidth]{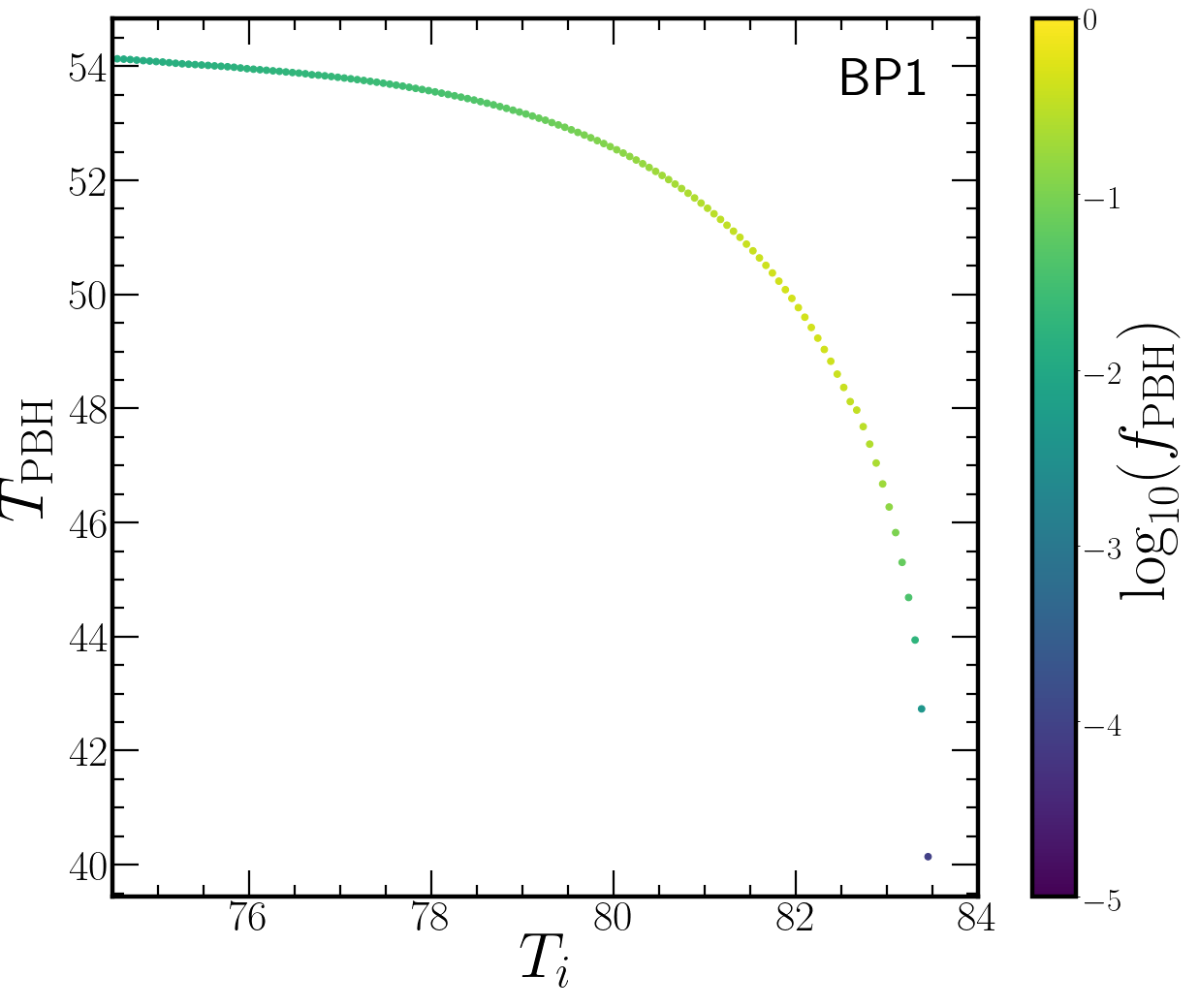}
    \caption{In the left panel, we show $(T_n-T_i)/T_n$ versus $(T_i-T_p)/T_p$ color-coded with $f_{\mathrm{PBH}}$. The points overproducing PBH, $f_{\rm PBH}>1$, are shown in orange and $f_{\rm PBH}$ in the range between $10^{-300}$ and $10^{-30}$ are shown in black. The values of $T_i$ are obtained maximizing $P(T_i)$ thereby $f_{\mathrm{PBH}}$. As $T_i$ approaches $T_p$, the value of $f_{\mathrm{PBH}}$ exponentially decreases. Conversely, when $T_i$ approaches $T_n$, PBHs are overproduced, $f_{\rm PBH}>1$. In the right panel, we present $T_i$ and the coressponding $T_{\mathrm{PBH}}$ color-coded with $f_{\mathrm{PBH}}$ for BP1. As $T_i$ increases, $f_{\mathrm{PBH}}$ reaches a maximum value and then decreases.}
    \label{fig:TiTpf}
\end{figure}

 For the numerical analysis, we determine $T_i$ by identifying the temperature that maximizes the PBH abundance $f_{\mathrm{PBH}}$~\autoref{eq:fpbh}, scanning over the range $T_p<T_i<T_n$~\cite{Baldes:2023rqv}.
The left panel of~\autoref{fig:TiTpf} justifies this scanning procedure. The orange points denotes PBH overproduction with $f_{\mathrm{PBH}}>1$ and black represent parameter points with $f_{\rm PBH}$ in the range between $10^{-300}$ and $10^{-30}$. For $T_i<T_p$, the value of $f_{\mathrm{PBH}}$ will be extremely small,
 $f_{\rm PBH}\ll 10^{-300}$,
and for $T_i>T_n$, the parameter points would be non-physical due to overproduction of PBH, $f_{\rm PBH}>1$.
In the right panel of~\autoref{fig:TiTpf}, we illustrate this behavior for BP1 by presenting $T_i$ and corresponding value of $T_{\mathrm{PBH}}$, color-coded with $f_{\mathrm{PBH}}$.
The value of $f_{\mathrm{PBH}}$ increases with $T_i$, and it reaches a threshold around $82.10\,\mathrm{GeV}$.
After this point, $f_{\mathrm{PBH}}$ drops due to a substantial decrease in $T_{\mathrm{PBH}}$.
Additional details on the PBH calculation, initial conditions for the evolution of energy densities, and  translation into temperature coordinates are provided in Appendix~\ref{app:PBH_calc}.

\section{Results}
\label{sec:Results}

The xSM model displays both single and multiple-step phase transitions. In this study, we comprehensively examine all possible phase transition patterns, with a focus on transitions resulting in PBH formation. In our parameter space analysis, we observe that $78.9\%$ of the parameter space points show PBH formation with a single-step phase transition.\footnote{See Appendix~\ref{app:paramscan} for more details on the parameter space scan.} This phase transition pattern is illustrated by the benchmark points BP1-BP3 in \autoref{tab:benchmark_points}. The remaining $21.1\%$ points enjoy a two-step phase transition.  BP4 in \autoref{tab:benchmark_points} is one such benchmark illustrating this scenario.

\begin{table}
\centering
\begin{tabular}{ c p{2.2cm}  p{2.2cm} p{2.2cm} p{2.2cm} p{2.2cm} p{2.2cm}  }
\toprule[1pt]
\toprule[1pt]
& BP1  &BP2 & BP3& BP4& BP5  \\
\cmidrule[0.5pt]{2-6}
$v_s$ [GeV]& $429.490$& $503.45$& $641.44$ &$1045.66$& $480.429$\\
$m_{h_2}$ [GeV]& $742.534$& $817.06$& $835.065$ & $1144.538$& $945.846$ \\
$\theta$& $-0.1597$& $-0.015458$ &$-0.15763$ & $-0.00208$& $-0.1548$ \\
$b_3$ [GeV]& $-2509.635$ &$-2348.49$& $-1990.700$& $-2130.053$& $-2272.226$\\
$b_4$ & $4.149478$ &$3.474019$& $2.28183$& $1.5696$& $3.4788$\\
\hline
$T_n$ [GeV]& $89.20$ &$101.82$ & $125.91$ &$190.61$& $117.98$  \\
$T_P$ [GeV]& $64.78$ &$75.71$ & $92.41$ & $142.46$& $114.09$  \\
$T_i$ [GeV]& $82.10$ &$92.58$ & $116.66$ & $177.45$& NA\\
$T_\mathrm{PBH}$ [GeV]& $59.53$ & $30.65$ & $71.30$ & $109.167$& NA  \\
$f^{\mathrm{EW}}_\mathrm{PBH}$& $0.457$&  $1.544\times 10^{-5}$ & $0.02263$ & $4.368 \times 10^{-5}$& NA  \\
$\alpha$& $6.31$& $4.525$ & $4.423$ & $3.65$& $0.645$\\
$R_{\star}\,[\mathrm{GeV}^{-1}]$& $4.52\times 10^{13}$ & $2.99\times 10^{13}$ & $2.95\times 10^{13}$ & $1.21\times 10^{13}$& $2.76\times 10^{12}$\\
\bottomrule[1pt]
\bottomrule[1pt]
\end{tabular}
\caption{Benchmark points for the singlet scalar model. BP1, BP2, and BP3 exemplify a single-step phase transition. BP4 is an example of two-step phase transition. BP5 does not lead to PBH formation, and is defined to highlight the importance of the U-shape profile presented in \autoref{BP1s3T}.}
\label{tab:benchmark_points}
\end{table}

\subsection{PBH formation in the xSM and shape of the effective potential}
\label{sec:PBH_xSM}

\begin{figure}[t!]
\includegraphics[width=0.55\textwidth]{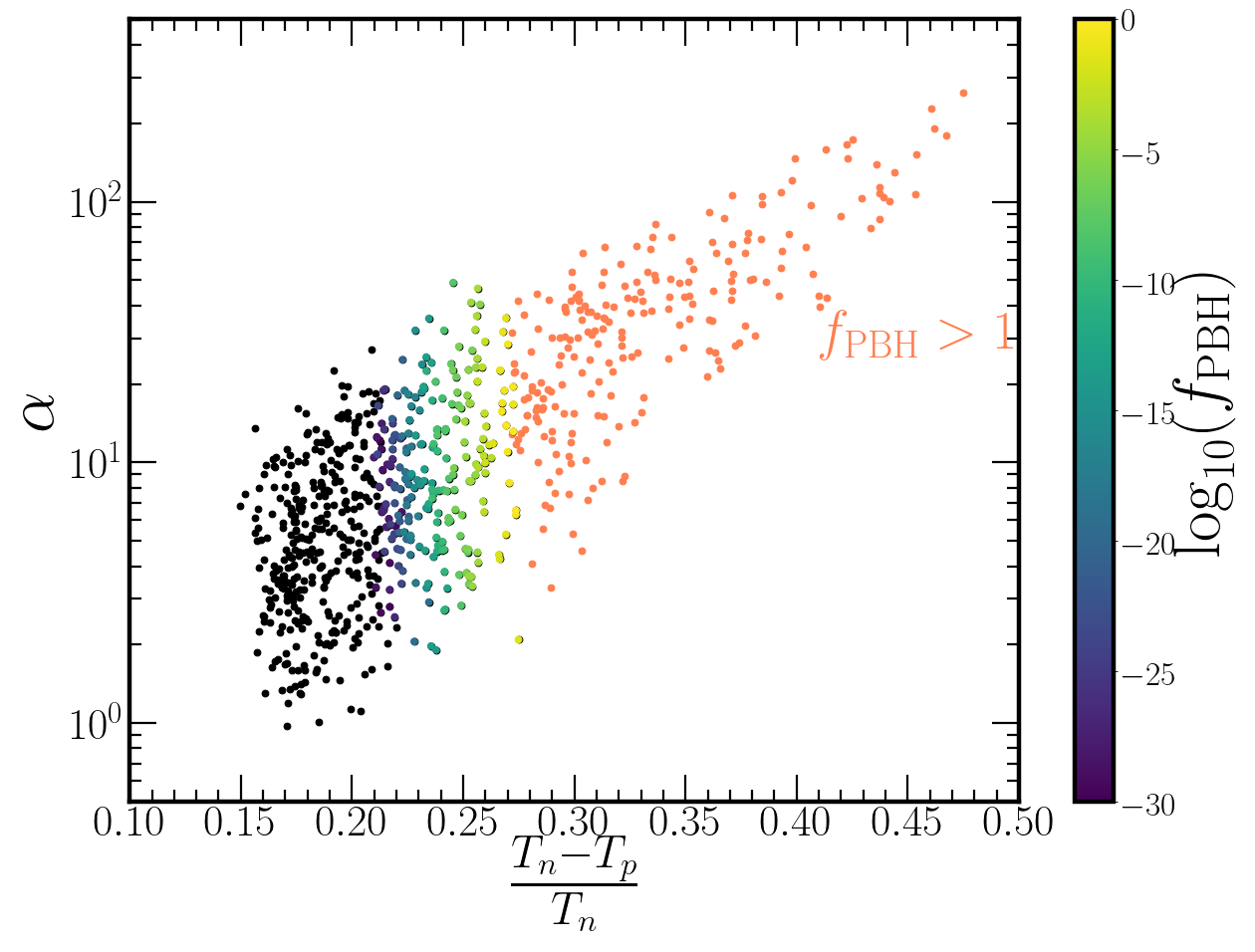}
\caption{We present the supercooling parameter ${(T_n - T_p)}/{T_n}$ plotted against $\alpha$ evaluated at $T_p$, with color coding indicating the value of $f_{\mathrm{PBH}}$. Parameter points exhibiting PBH overproduction with $f_{\mathrm{PBH}} > 1$ are represented in orange and $f_{\rm PBH}$ in the range between $10^{-300}$ and $10^{-30}$ are shown in black.}
\label{supercooling}
\end{figure}

In this section, we examine the profile required in the effective potential of the xSM model that triggers the formation of PBHs.  In~\autoref{supercooling}, we show $(T_n-T_p)/{T_n}$, which quantifies the extent of supercooling versus $\alpha$ evaluated at $T_p$ for the parameter points that show PBH formation.
It is clear that PBH forms only if we have sufficient supercooling $(T_n-T_p)/T_n\gtrsim0.12$. However, if the supercooling is too large $(T_n-T_p)/T_n\gtrsim0.3$, the PBH is overproduced and can overclose the Universe. Within the viable range, the amount of PBH formed is highly sensitive to the extent of supercooling with only a minor dependence on $\alpha$, as supercooling exponentially affects the probability of PBH formation.

\begin{figure}[t!]
\includegraphics[height=0.28\textwidth]{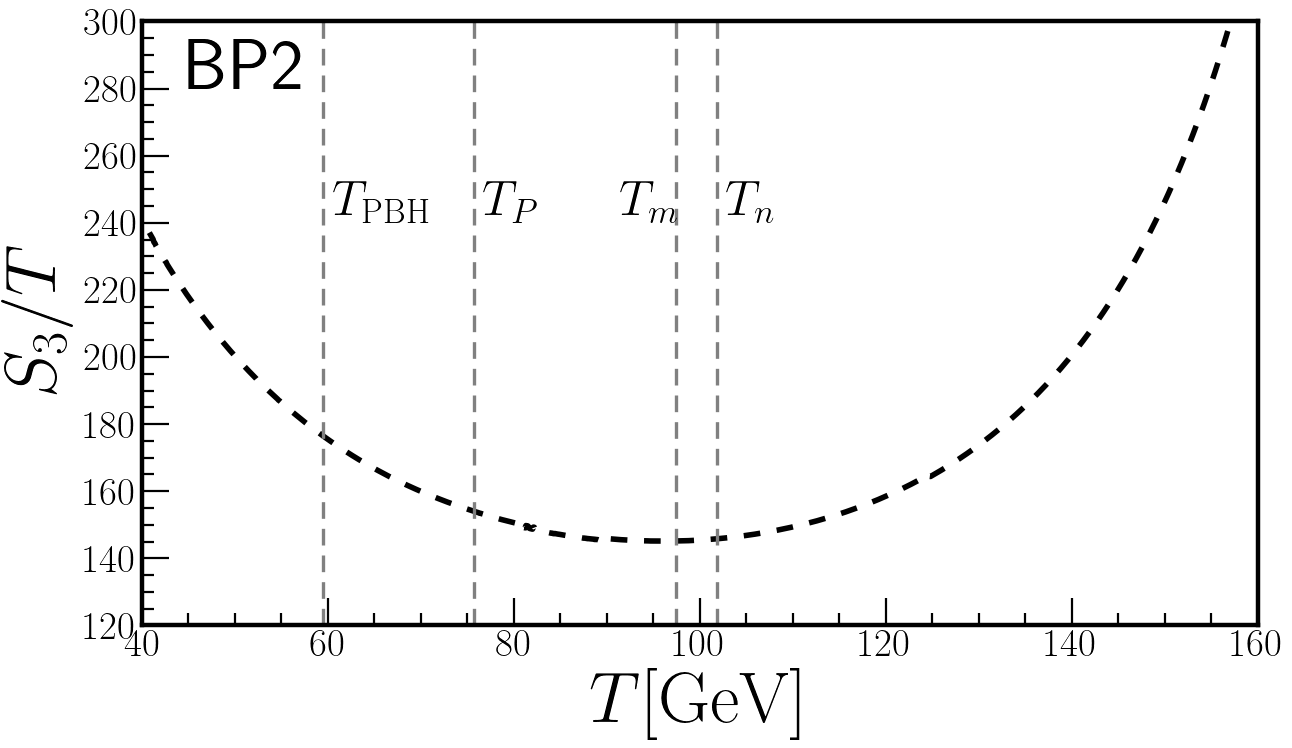}
\includegraphics[height=0.28\textwidth]{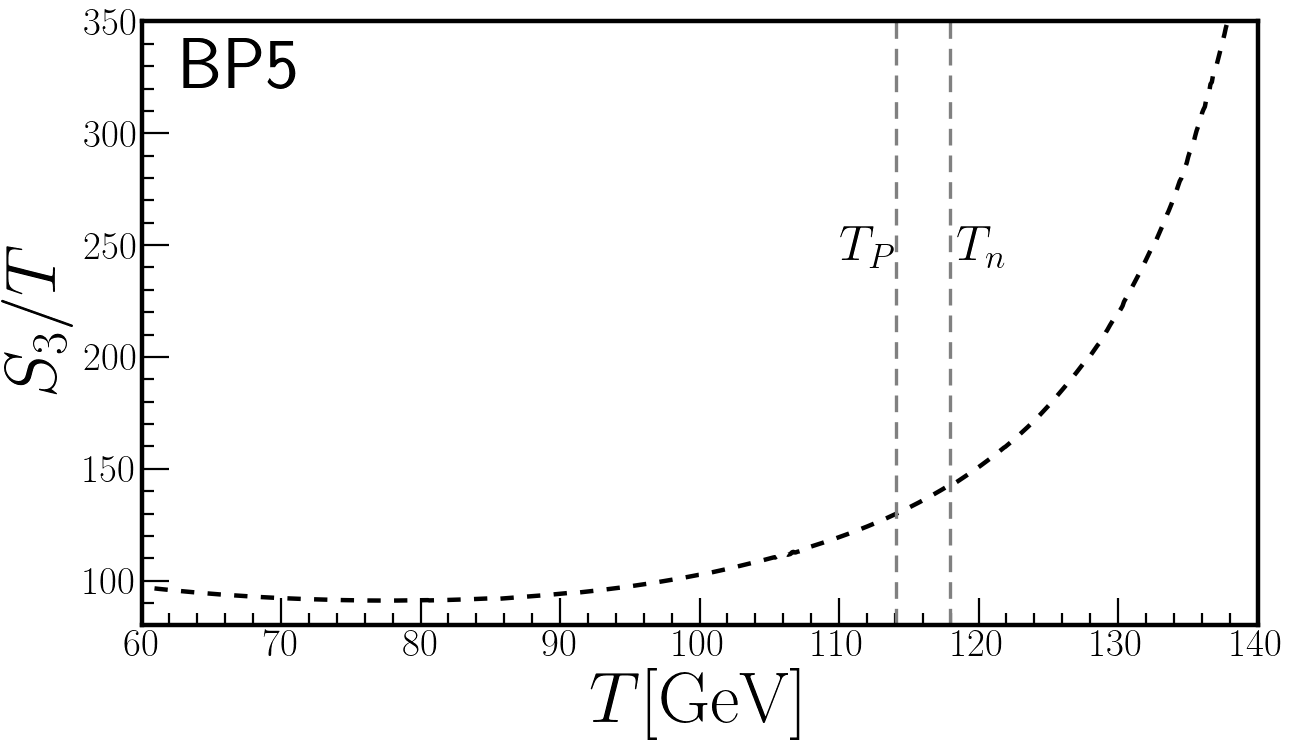}
\caption{The evolution of ${S_3}/{T}$ with temperature for the BP2 (left panel) and BP5 (right panel), as defined in~\autoref{tab:benchmark_points}.  BP2 exhibits a U-shaped behavior of $S_3/T$ near the nucleation temperature, whereas BP5 does not show this profile. BP1 undergoes significant supercooling, resulting in PBH formation. In contrast, BP5 experiences minimal supercooling and does not lead to PBH formation. The nucleation temperature $T_n$ and percolation temperature $T_p$ are depicted with dotted lines. Furthermore, for BP2, we have indicated the PBH formation temperature $T_{\mathrm{PBH}}$ and the temperature where the tunneling rate is maximum $T_m$.}
\label{BP1s3T}
\end{figure}

To investigate the primary factors that trigger supercooling, we analyze the evolution of $S_3/T$ as a function of temperature for two benchmark points: BP2 and BP5. Details regarding these BPs are provided in~\autoref{tab:benchmark_points}. The left panel of~\autoref{BP1s3T} illustrates the evolution of $S_3/T$ with temperature for BP2, which exhibits PBH formation. Notably, $S_3/T$ displays a U-shaped behavior with temperature, leading to an exponential suppression of the tunneling rate at both high and low temperatures~\cite{Alves:2019igs}. This suppression enables significant bubble formation only within a specific intermediate temperature range. In this scenario, the tunneling rate can be approximated by~\cite{Athron:2022mmm,Athron:2023rfq,Kanemura:2024pae}
\begin{equation}
    \label{equ:tunneling_rate}
    \Gamma=\Gamma_0e^{-\frac{1}{2}\beta_V(T-T_m)^2},
\end{equation}
where $T_m$ is the temperature at which tunneling rate is maximum and $\beta_V\equiv\left .\frac {d^2(S_3/T)}{dT^2}  \right |_{T=T_m}$. This characteristic behavior of $S_3/T$ and subsequently of the tunneling rate leads to a supercooling phase transition, which helps the production of PBHs. In contrast, the right panel of~\autoref{BP1s3T} shows the temperature evolution of $S_3/T$ for BP5, which does not present significant supercooling and does not result in PBH formation. BP5 lacks the U-shaped behavior of $S_3/T$ near the nucleation temperature. Hence, the tunneling rate increases steadily between the temperatures $T_n$ and $T_p$, leading to bubble percolation shortly after the nucleation of the first bubble.
\begin{figure}[tb!]
    \includegraphics[height=0.4\textwidth]{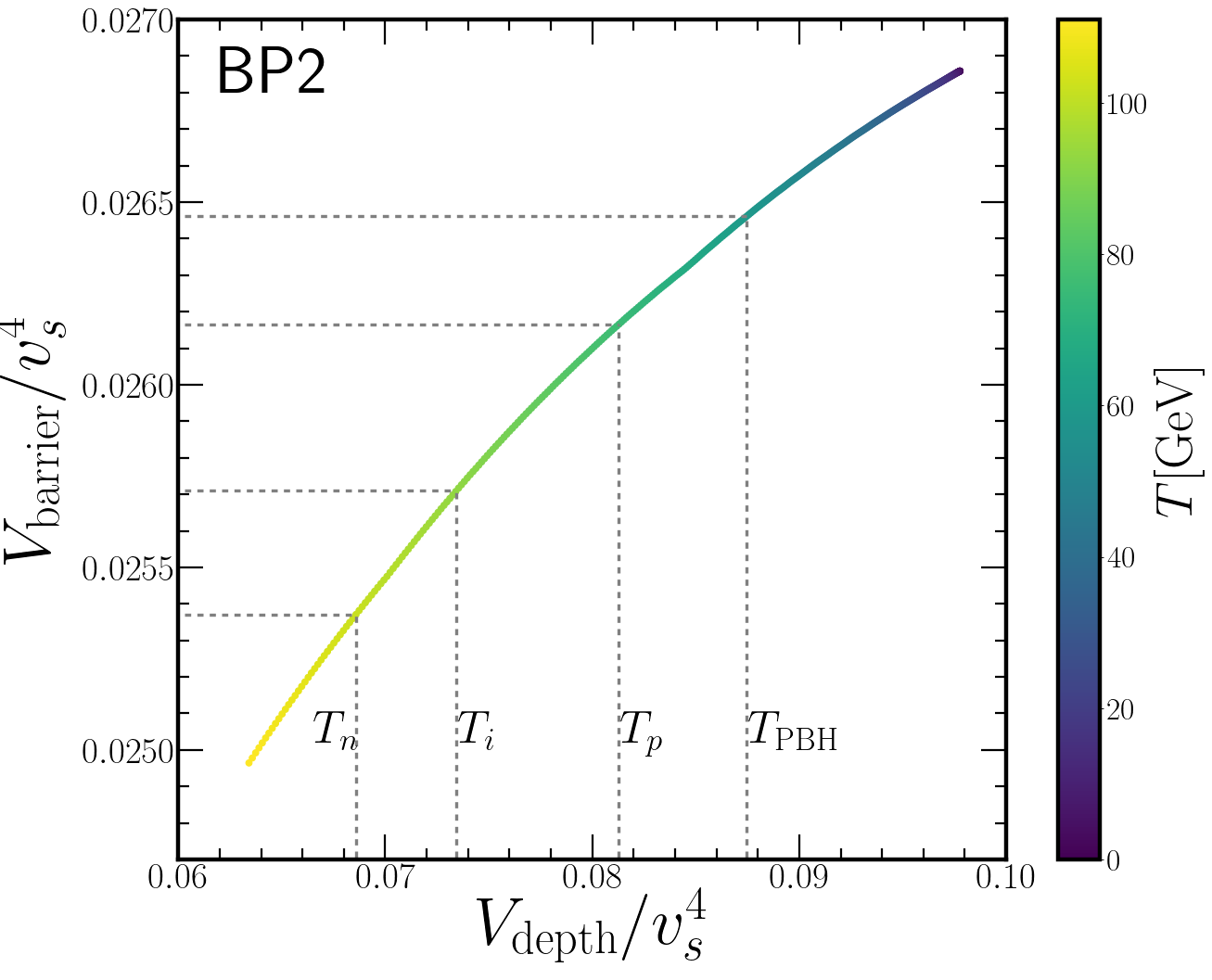}
    \includegraphics[height=0.4\textwidth]{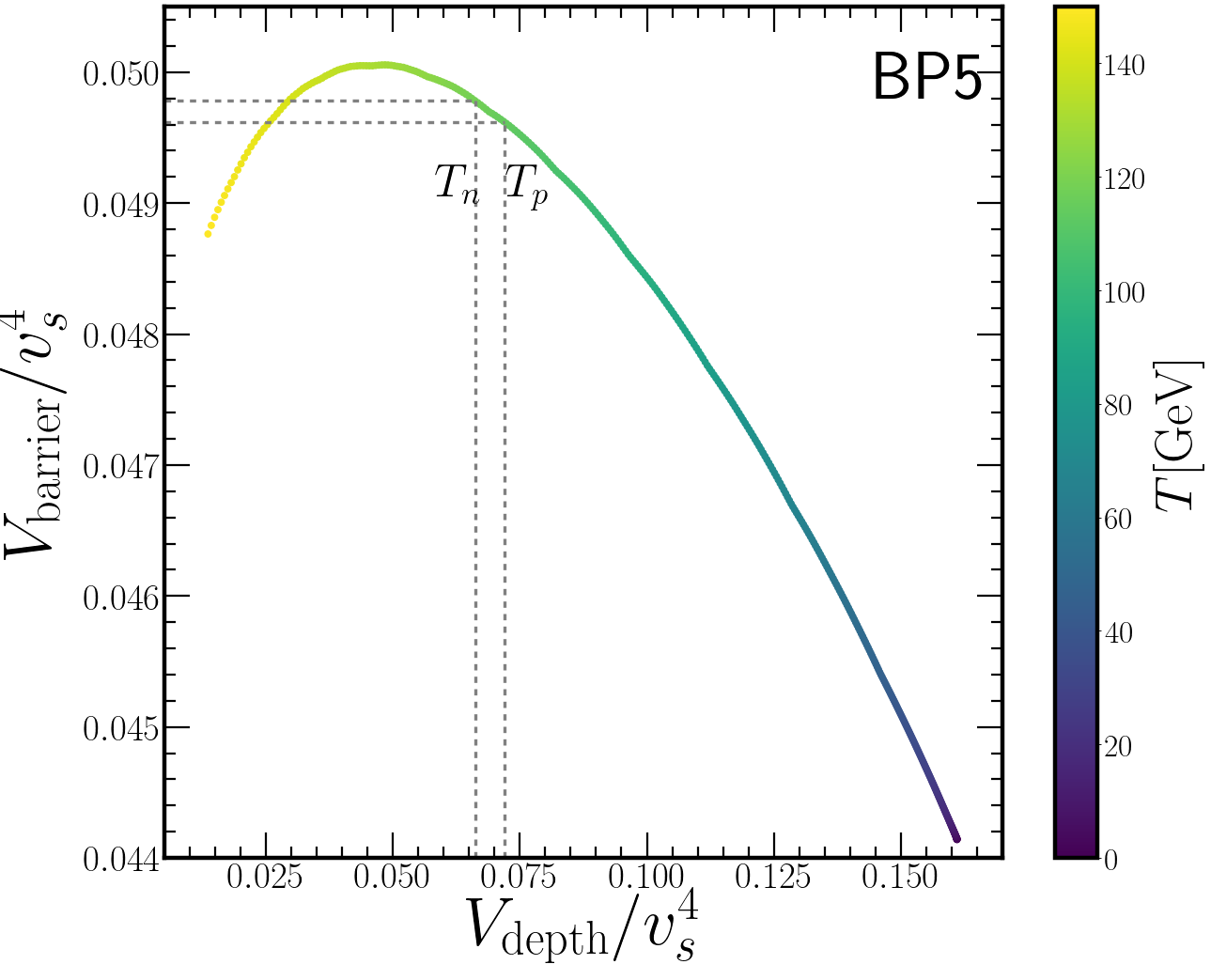}
    \caption{We show the thermal evolution of the shape of the potential by plotting $V_{\mathrm{barrier}}$ versus $V_{\mathrm{depth}}$ color-coded with temperature for the BP2 (left panel) and BP5 (right panel) provided in~\autoref{tab:benchmark_points}. The benchmark point BP2  show U-shaped behavior for $S_3/T$ near nucleation temperature, while BP5 does not show such behavior. For BP1, the height of the barrier increases with temperature, while for BP5, the height decreases with temperature, exhibiting a distinctive behavior.}
    \label{fig:BP_Veff_T1}
\end{figure}

In some model-agnostic studies of PBH formation from first-order phase transition~\cite{Liu:2021svg, Kawana:2022olo}, the tunneling rate, given by~\autoref{eq:Gamma}, is often approximated by $\Gamma(t)\approx \Gamma_0e^{\beta(t-t_n)}$, where $\Gamma_0$ is the value of tunneling rate at the nucleation time $t_n$ and $\beta\equiv-{dS_3(t)}/{dt}|_{t=t_n}$. However, it is evident from the~\autoref{BP1s3T} that such an approximation may not consistently hold for the xSM model and other models which have a significant zero-temperature barrier.
This is because the U-shaped behavior of $S_3/T$ near $T_n$ is not captured by this simple exponential approximation. Thus, it is more appropriate to use the tunneling rate in~\autoref{equ:tunneling_rate}, when $S_3/T$ exhibits this U-shaped behavior around $T_n$.

 To gain further insight into the scenarios that could result in the U-shaped behavior of $S_3/T$ with temperature, we scrutinize the shape of the potential at finite temperature, focusing on barrier formation and vacuum upliftment~\cite{Goncalves:2021egx,Goncalves:2022wbp,Goncalves:2023svb}. In~\autoref{fig:BP_Veff_T1}, we illustrate the thermal evolution of the barrier and the depth of the true vacuum for the BP2. For simplicity, the barrier is calculated along the line that directly connects the false and true vacuum. As the temperature cools down, both depth and barrier increase. The increase in barrier height makes tunneling from the false to true vacuum more challenging; as a consequence, we observe the U-shaped behavior for $S_{3}/T$ behavior for BP2, as indicated in~\autoref{BP1s3T}. The $83.2\%$ of the parameter points that lead to PBH have this behavior where the barrier height increases between $T_n$ and $T_p$, which serves as a sufficient condition for the U-shaped behavior of $S_3/T$ with temperature.

\begin{figure}[t!]
\includegraphics[height=0.265\textwidth]{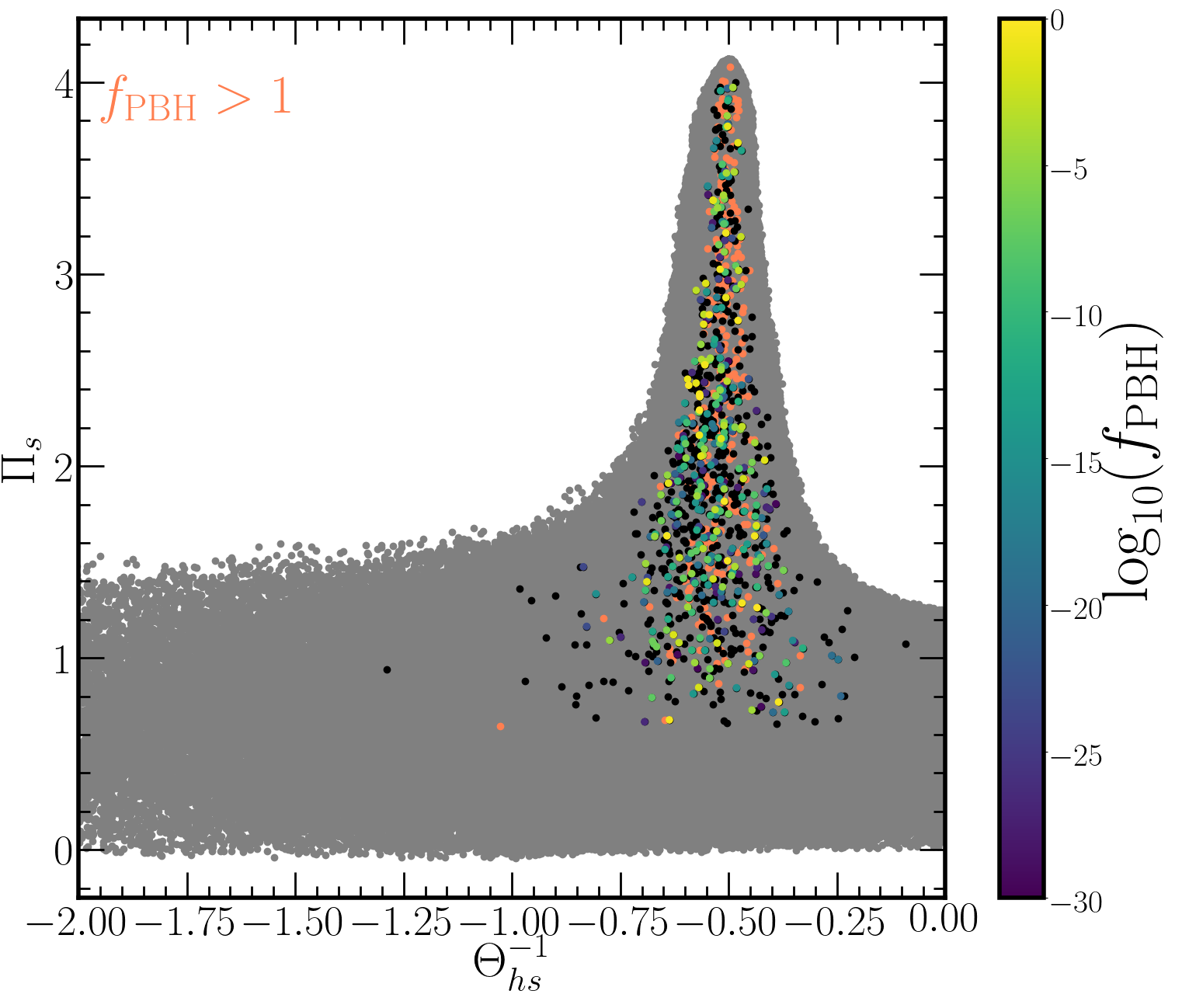}
\includegraphics[height=0.265\textwidth]{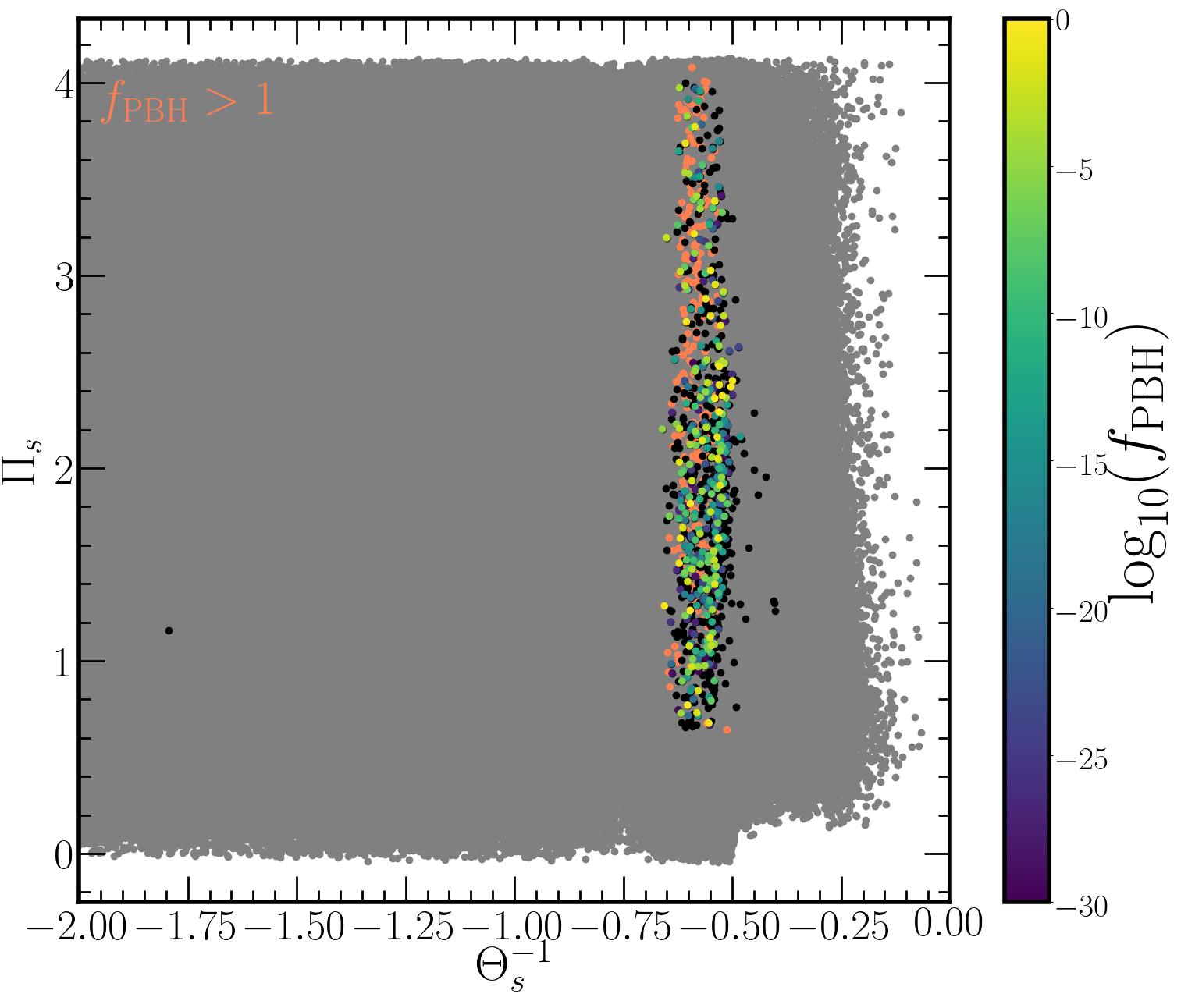}
\includegraphics[height=0.265\textwidth]{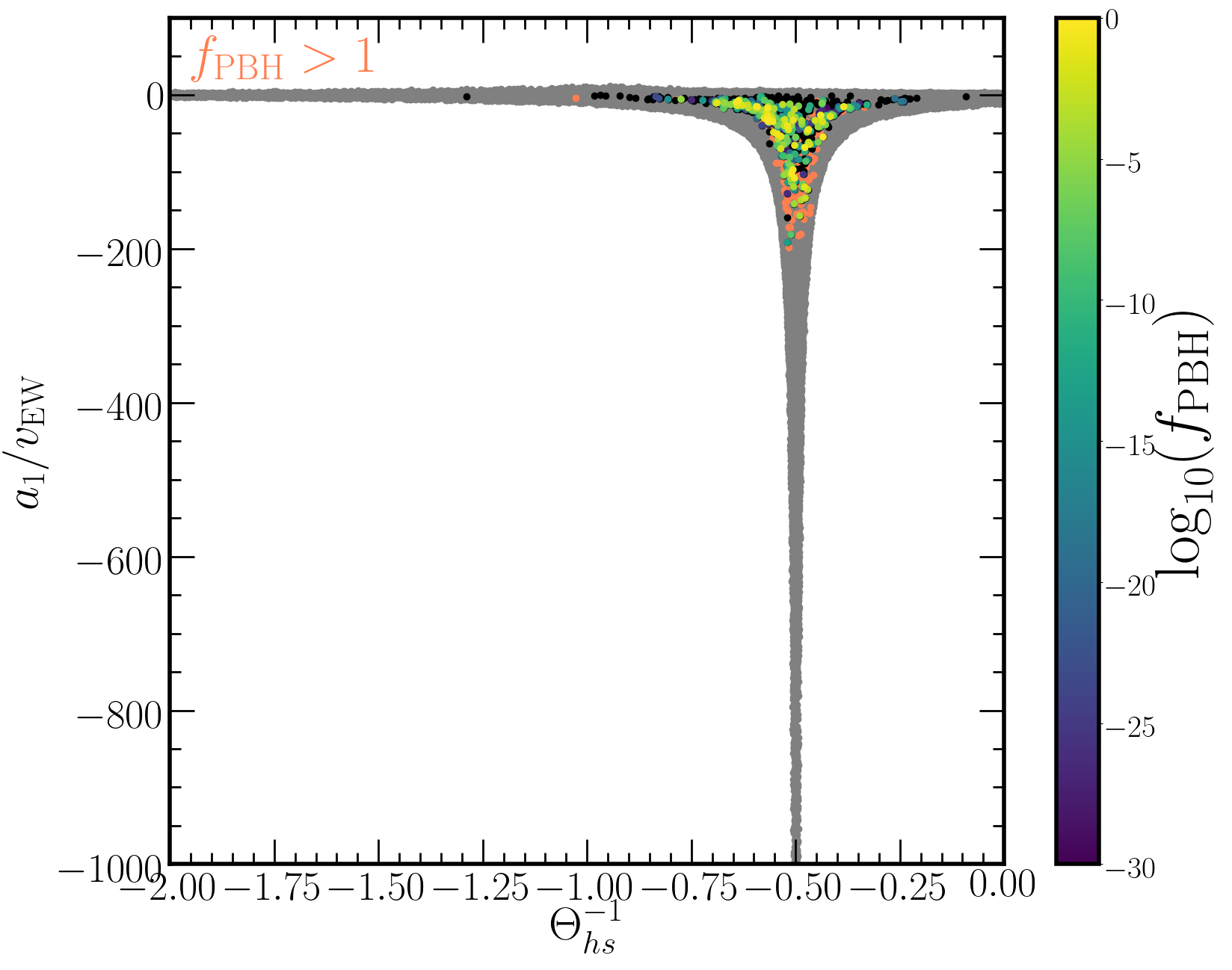}
\caption{We show the inverse of ratios $\Theta_{hs}$ and $\Theta_{s}$  versus thermal mass $\Pi_s$ color coded by $\log_{10}(f_{\mathrm{PBH}})$ for the center and right panel respectively. In the right panel we show the inverse of $\Theta_{hs}$ versus $a_1$ normalized by $v_{\mathrm{EW}}$. Parameter points satisfying theoretical and experimental constraints are denoted in gray, whereas those exhibiting PBH formation with $f_{\rm PBH}$ in the range between $10^{-300}$ and $10^{-30}$ are highlighted in black and orange color denotes the parameter excluded due to overproduction of DM $f_{\mathrm{PBH}}>1$. Parameter points that lead to PBH formation have thermal mass $\Pi_s \gtrsim 0.8$, negative cubic couplings, and the values for $\Theta_{hs}^{-1}$ and $\Theta_{s}^{-1}$ are roughly in the vicinity of $-0.5$ and $-0.57$.}
\label{Frac_Pis_PBH}
\end{figure}

From the observations in~\autoref{fig:BP_Veff_T1}, it becomes apparent that the tree-level barrier plays a crucial role in the formation of PBHs in the xSM. The cubic terms in the~\autoref{eq:V0} provide the dominant contribution to the barrier at zero temperature. We define the cubic terms normalized by the quartic terms as
\begin{equation}
\Theta_{hs}\equiv \frac {a_1}{a_2 v_s}, \quad \quad \Theta_{s}\equiv\frac {4 b_{3}}{3 b_4 v_s}.
\label{eq:Theta_def}
\end{equation}
In the left (center) panel of the~\autoref{Frac_Pis_PBH}, we show  $\Theta_{hs}^{-1}$ ($\Theta_s^{-1}$) versus thermal mass $\Pi_s$ color-coded by the $\log_{10}(f_{\mathrm{PBH}})$. In the right panel, we show the cubic term $a_1$ normalized by the $v_{\mathrm{EW}}$ versus $\Theta_{hs}^{-1}$ color coded by the $\log_{10}(f_{\mathrm{PBH}})$. The gray color denotes the parameter points that satisfy theoretical and experimental constraints, while those displaying PBH formation follow the color code associated with $\log_{10}(f_{\mathrm{PBH}})$. The PBHs favor the parameter region with thermal mass $\Pi_s\gtrsim 0.8$, while the normalized cubic terms have relatively narrow allowed ranges: $\Theta_{hs}^{-1}\simeq -0.5$ and $\Theta_s^{-1}\simeq-0.57$. The PBH formation requires the cubic term to be negative and the corresponding quartic term to be positive.

\begin{figure}
    \includegraphics[height=0.38\textwidth]{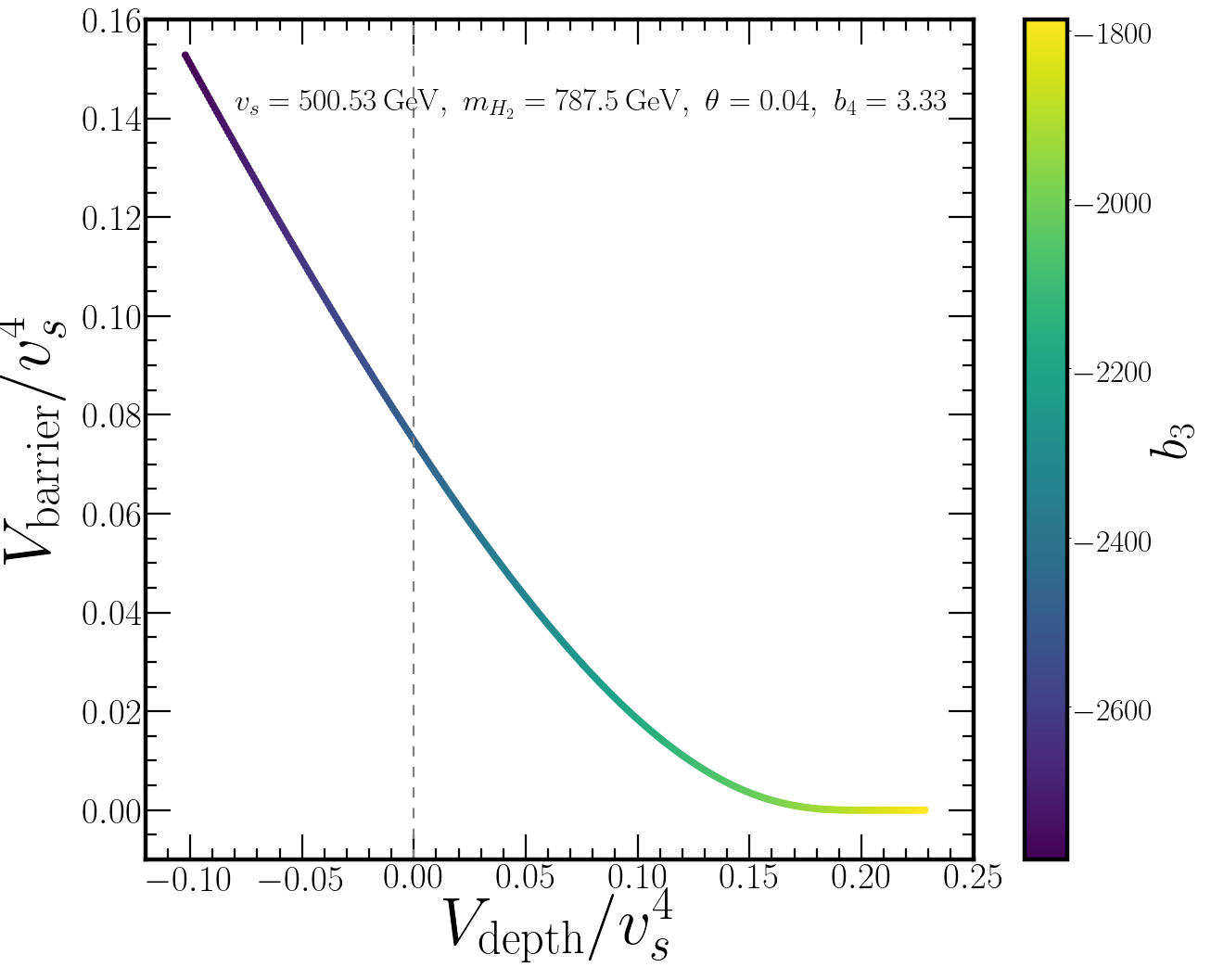}
    \includegraphics[height=0.38\textwidth]{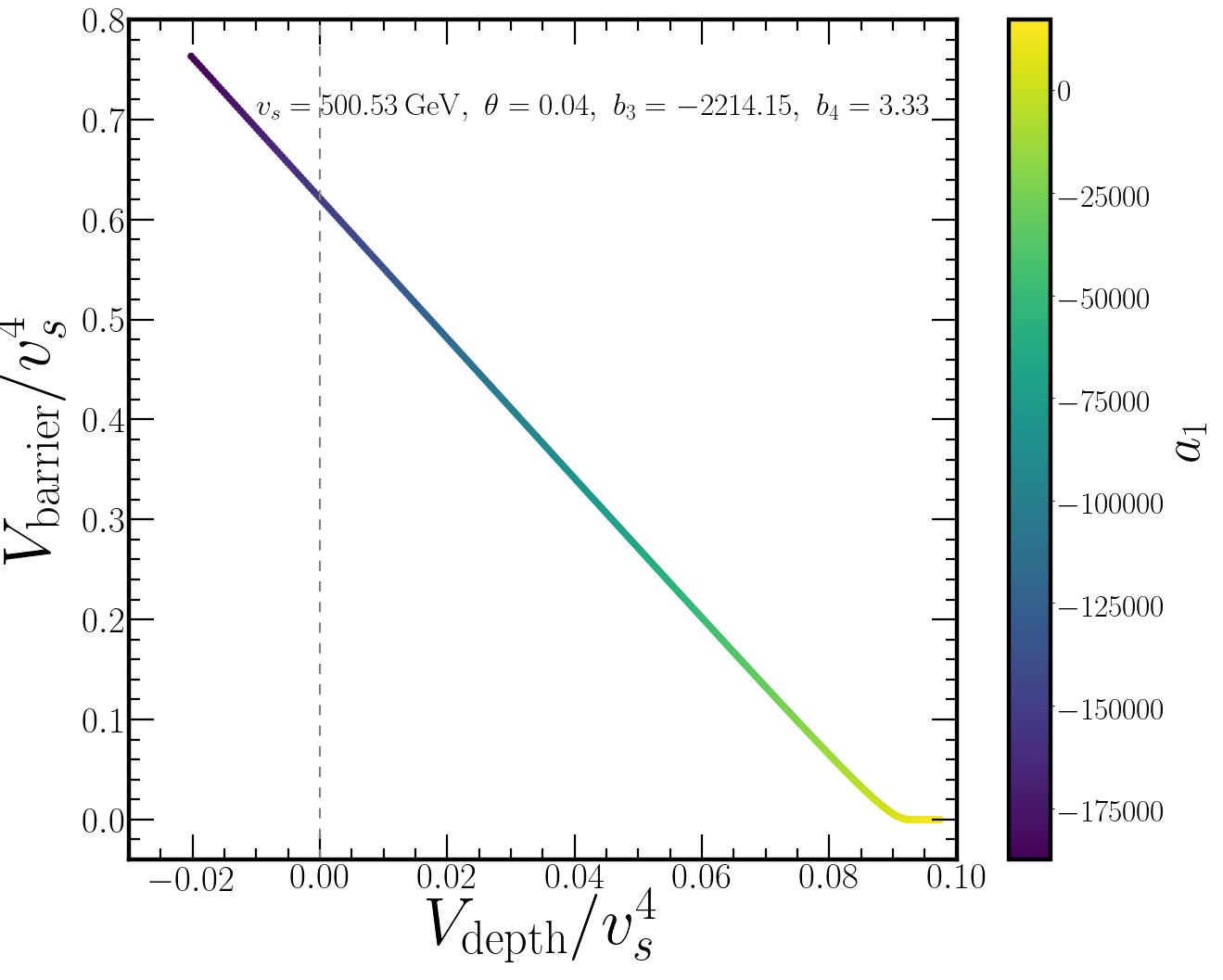}
    \caption{We show the dependence of cubic terms $b_3$ (left panel) and $a_1$ (right panel) on the shape of potential at zero temperature by plotting $V_{\mathrm{barrier}}$ versus $V_{\mathrm{depth}}$ for different values of $b_3$ and $a_1$. For the left panel, we fixed $v_s=500.53\,\mathrm{GeV},\,\, m_{h_2}=787.5\,\mathrm{GeV},\,\,\theta=0.04$, and $b_4=3.33$. As the values of $b_3$ increase, the height of the barrier decreases, and eventually, the barrier disappears. For the large negative value of $b_3$,  the EW vacuum stability can be violated, as the depth reaches negative values. For the right panel we fixed $v_s=500.53\,\mathrm{GeV},\,\,\theta=0.04,\,\,b_3=-2214.15$ and $b_4=3.33$. A similar trend emerges for $a_1$: as cubic terms grow more negative, the barrier height increases while the depth decreases.}
    \label{fig:Veff_T_Theta}
\end{figure}
This sensitive dependence of the PBH formation on $\Theta_{hs}$ and $\Theta_{s}$ can also be related to the influence of the cubic terms on the barrier and depth of the effective potential. To illustrate it, in the left panel of~\autoref{fig:Veff_T_Theta}, we examine the shape of the potential at zero temperature by fixing $v_s=500.53\,\mathrm{GeV},\,\, m_{h_2}=787.5\,\mathrm{GeV},\,\,\theta=0.04,$ and $b_4=3.33$, while varying $b_3$, which is equivalent to varying $\Theta_s$. The height of the barrier and depth of the EW broken vacuum are inversely correlated. As $b_3$ becomes increasingly negative, the cubic term dominates over the quartic term, the EW broken vacuum is uplifted compared to the EW symmetric vacuum and thereby, its depth decreases. On the other hand, the domination of the cubic term leads to the increase in the barrier height. As the value of $b_3$ becomes more negative, the depth decreases, and the height of the barrier increases. Around $b_3\approx-2474$ ($\Theta_s^{-1}\approx-0.506$), depth becomes negative and the stability of the EW vacuum is violated. Hence, in view of the values of $\Theta_s^{-1}$ leading to PBH formation shown in \autoref{Frac_Pis_PBH},  PBH generation favors parameter space where the potential has a shallow EW vacuum and sufficiently high barrier. A similar pattern can also be found for $a_1$ in the right panel of~\autoref{fig:Veff_T_Theta}; as cubic terms become more negative, the barrier increases, and depth decreases. In the majority of the parameter space conducive to PBH formation, it is observed that both cubic couplings, $b_3$ and $a_1$, collectively contribute to the generation of the barrier.

\subsection{Microlensing constraints}
\label{sec:microlensing}

\begin{figure}[!tb]
    \centering
    \includegraphics[width=0.5\textwidth]{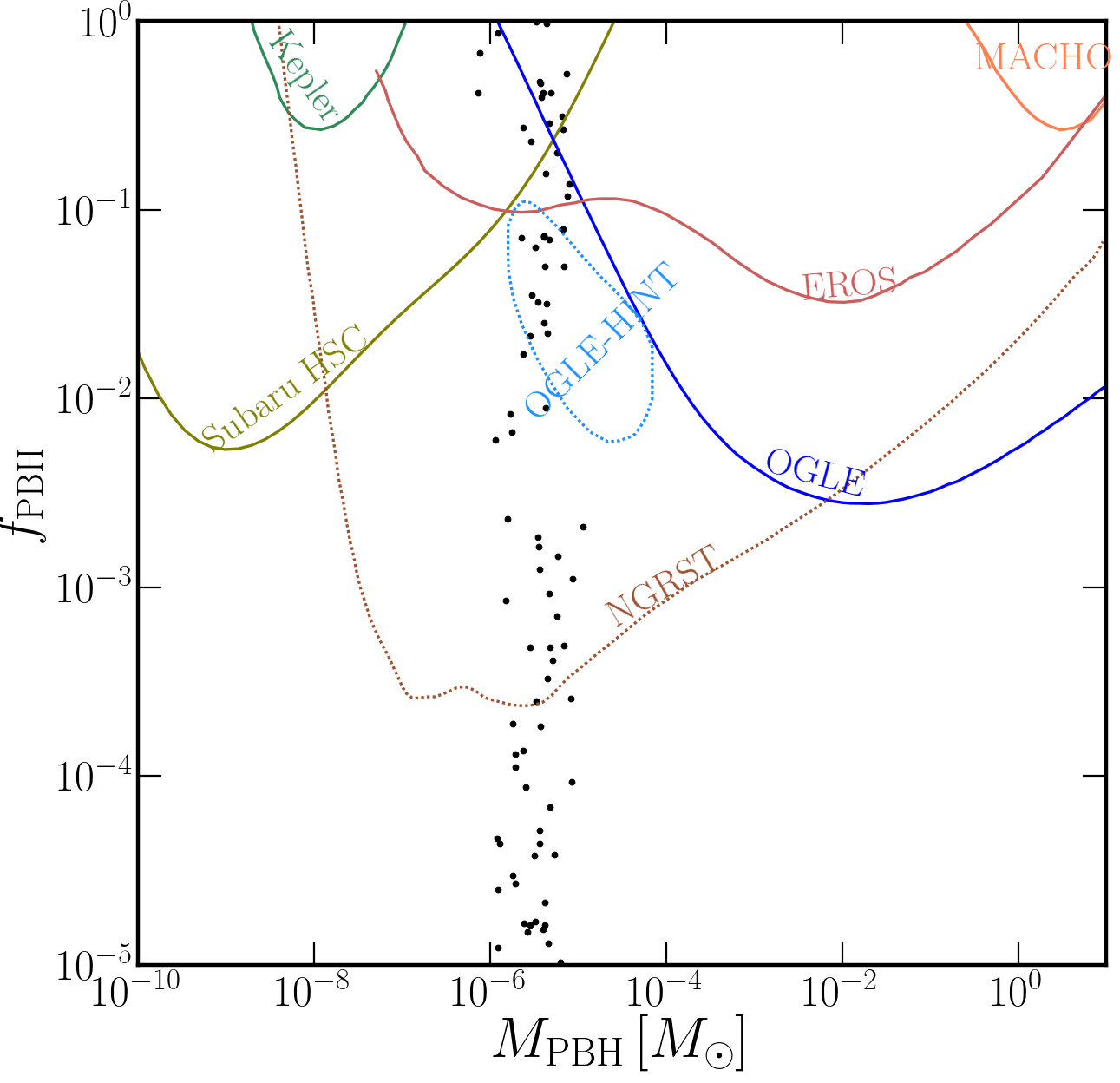}
    \caption{Fraction of PBH contribution to dark matter density today $f_{\mathrm{PBH}}\equiv \rho_{\mathrm{PBH}}/\rho_{\mathrm{DM}}$ as a function of the PBH mass $M_\text{PBH}[M_{\odot}]$  in the xSM model, where $M_{\odot}$ is the solar mass. In addition, we present the microlensing constraints from Kepler~\cite{Griest:2013aaa}, MACHO~\cite{Macho:2000nvd}, EROS~\cite{EROS-2:2006ryy}, SUBARU-HSC~\cite{ Croon:2020ouk, Niikura:2017zjd}, and OGLE~\cite{Mroz:2024mse, Mroz:2024wag,2017Natur.548..183M, Niikura:2019kqi}. We also include the projected sensitivity of NGRST~\cite{DeRocco:2023gde} and early hints of observation of Earth-sized PBHs by OGLE~\cite{Niikura:2019kqi}, indicated by dotted lines.}
    \label{fig:PBH}
\end{figure}

The gravitational effects of the PBHs can be observed through its microlensing effects. In~\autoref{fig:PBH},
we present the fraction of the contribution of PBHs to dark matter density today, $f_\mathrm{PBH}$, as a function of the PBH mass derived from the xSM framework together with microlensing constraints from several experiments.\footnote{The microlensing constraints for the PBH are obtained from {\tt PBHbounds} at \url{https://github.com/bradkav/PBHbounds}.} In~\autoref{fig:PBH}, the current  microlensing constraints from Kepler~\cite{Griest:2013aaa}, MACHO~\cite{Macho:2000nvd}, EROS~\cite{EROS-2:2006ryy,Blaineau:2022nhy}, SUBARU-HSC~\cite{ Croon:2020ouk, Niikura:2017zjd}, and OGLE~\cite{Mroz:2024mse, Mroz:2024wag, 2017Natur.548..183M,Niikura:2019kqi} are depicted with solid lines. Some comments are in order. First, the mass distribution of PBHs is remarkably narrow, approximately $10^{-5}$ solar masses. This narrow distribution is closely related to the temperature at which these PBHs form, as evidenced by~\autoref{equ:MPBH},
which aligns with the formation of PBH from first-order phase transition at the electroweak scale. Second, while the model in principle allows for the generation of significant dark matter fractions from PBHs, potentially reaching nearly 100\% of the dark matter density, the large $f_\text{PBH}$ regions are constrained by microlensing experiments. This implies that the contribution of PBHs to the dark matter density in this model can be as high as $f_\text{PBH}\approx 10^{-1}$, with  OGLE, Subaru-HSC, and Eros experiments placing the most stringent limits~\cite{2017Natur.548..183M,Niikura:2019kqi,PhysRevLett.111.181302,Niikura:2017zjd,EROS-2:2006ryy}.

The OGLE experiment, through its 5-year observations of stars in the Galactic bulge, detected six ultrashort timescale events,  hinting at the possible presence of Earth-sized PBH in the outer solar system~\cite{2017Natur.548..183M,Niikura:2019kqi,Scholtz:2019csj}. Alternatively, these events could also be linked to the new population of free-floating planets, one of which may have been captured by the solar system. Within the parameter space of the extended xSM, PBH formation is feasible, aligning well with the OGLE observations; for instance, benchmark point BP3 in \autoref{tab:benchmark_points} exemplifies this correspondence. This PBH parameter space could be explained using PBH produced from inflation~\cite{Yi:2023npi, Cai:2023uhc, Fu:2022ypp, Solbi:2021rse, Motohashi:2019rhu, Fu:2019ttf}, the collapse of domain wall~\cite{Ge:2023rrq} and axion star~\cite{Sugiyama:2021xqg}, but this is the first time, as per our knowledge, the OGLE-hint can be attributed to electroweak phase transition in a realistic BSM model that generates PBHs from delayed vacuum transition.

The forthcoming Nancy Grace Roman Space Telescope (NGRST), scheduled for launch in 2027, offers promising prospects for further investigating the PBH hypothesis concerning the six ultrashort timescale events detected by OGLE~\cite{DeRocco:2023gde}. In~\autoref{fig:PBH}, we present the sensitivity of NGRST and OGLE-hint with dotted lines. The NGRST has the sensitivity to probe PBH with a dark matter fraction as low as $10^{-4}$.
Confirmation of the PBH explanation for ultrashort timescale events by these experiments could lead us to the favored parameter space in BSM theories that undergo EWPT with significant supercooling. Alternatively, if this scenario is ruled out, the high sensitivity of NGRST and other microlensing experiments will impose stringent constraints not only on the parameter space of the xSM model but also on similar models that induce PBH formation through delayed EW phase transitions.

\subsection{Gravitational wave experiments}
\label{sec:GW}

\begin{figure}[!b]
    \centering
    \includegraphics[width=0.55\textwidth]{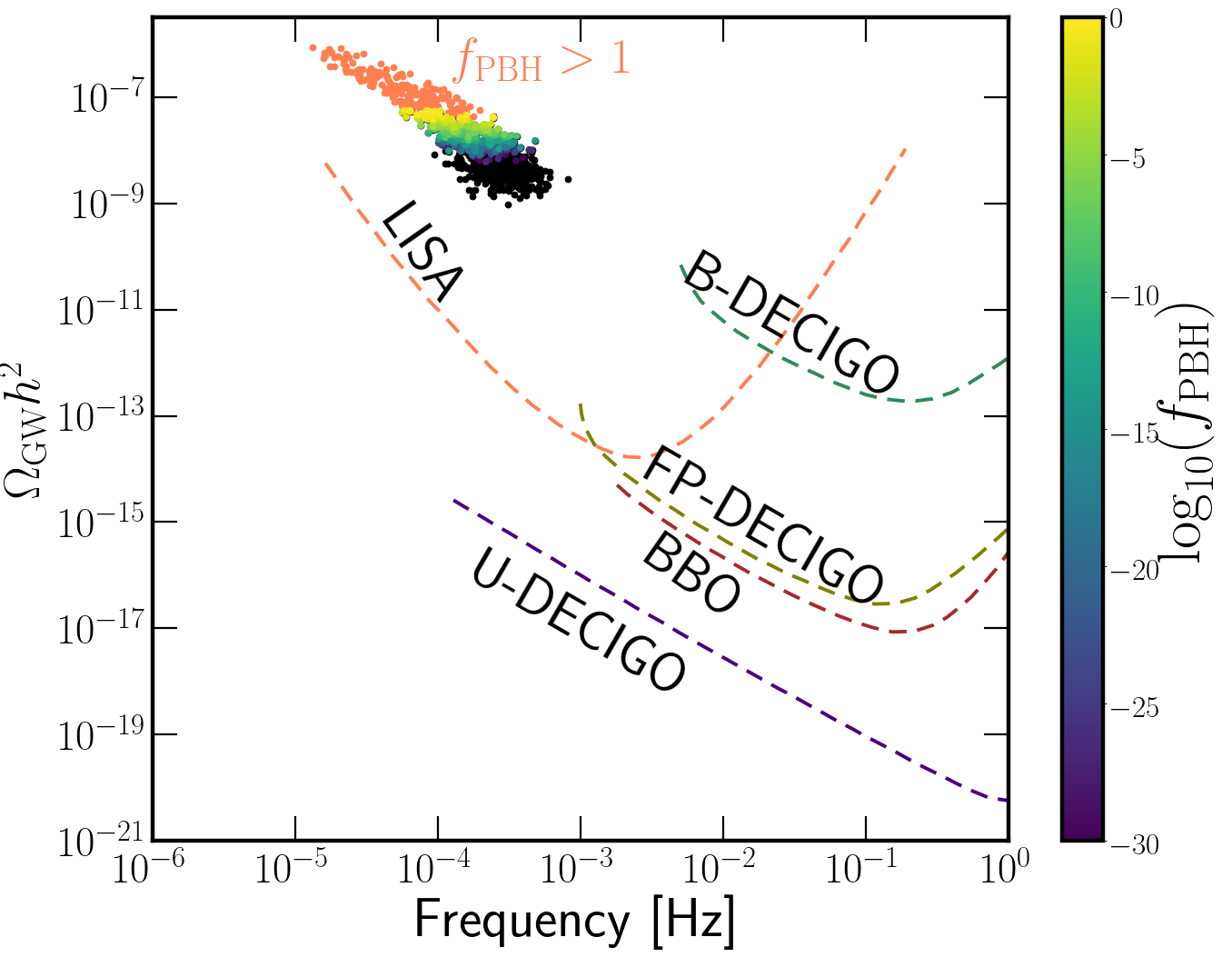}
    \caption{The maximum amplitude of the gravitational wave spectrum as a function of peak frequency color coded by  $f_{\mathrm{PBH}}$ and the orange color denotes the parameter excluded due to overproduction of DM $f_{\mathrm{PBH}}>1$ whereas $f_{\rm PBH}$ in the range between $10^{-300}$ and $10^{-30}$ are shown in black. As discussed in \autoref{sec:GW}, the requirement that at least one PBH forms in the observable Universe imposes a constraint of $f_\mathrm{PBH} > 10^{-30}$.
    We also present the experimental sensitivities for LISA~\cite{Caprini:2015zlo}, BBO~\cite{Corbin:2005ny}, and three stages of DECIGO~\cite{Kudoh:2005as}. The sensitivity from LISA was evaluated considering the duration of the mission of 5 years~\cite{Caprini:2015zlo}.
    }
    \label{fig:GW_few}
\end{figure}

In the region favoring PBH formation, the large energy release during the phase transition results in detectable gravitational wave signals. For more details on the GW calculation, see the end of~\autoref{sec:EWPT} and~Appendix~\ref{app:GW}. In~\autoref{fig:GW_few}, we show the maximum amplitude of gravitational waves produced during first-order phase transition as a function of its peak frequency, color coded by the PBH fraction $f_{\mathrm{PBH}}$.  Additionally, we  provide the projected sensitivity curves for LISA~\cite{Caprini:2015zlo}, BBO~\cite{Corbin:2005ny} and three stages of  DECIGO~\cite{Kudoh:2005as}  GW experiments.  PBH formation from first-order phase transition requires supercooling and a large value of $\alpha$, which coincides with promising gravitational wave signatures. Since the phase transition occurs at the electroweak scale, it naturally falls within the frequency range detectable by LISA. We find that all parameter points demonstrating PBH formation  with $f_{\mathrm{PBH}}>10^{-300}$ in the xSM model can be probed by LISA.

To obtain relevant phenomenology, we will need a minimum value of $f_{\rm PBH}$ that allows for the formation of at least one PBH in the observable Universe. Using the critical density $\rho_c\simeq9\times10^{-27}~\mathrm{kg/m^3}$~\cite{pdg}, the radius of observable Universe $r_h\simeq4.4\times10^{26}~\mathrm{m}$, and  considering a single PBH with mass around $10^{-5} M_{\odot}$ (see Fig.~\ref{fig:PBH}), we obtain the minimum threshold for the PBH fraction as $f_{\rm PBH}\sim 10^{-30}$. Since LISA is sensitive across the entire region $10^{-30}<f_{\rm PBH}<1$, it has the potential to probe the parameter space  that gives rise to PBHs in the xSM within the observable Universe, either confirming our PBH scenario or completely ruling it out.

\subsection{Collider constraints}
\begin{figure}[!t]
    \includegraphics[height=0.38\textwidth]{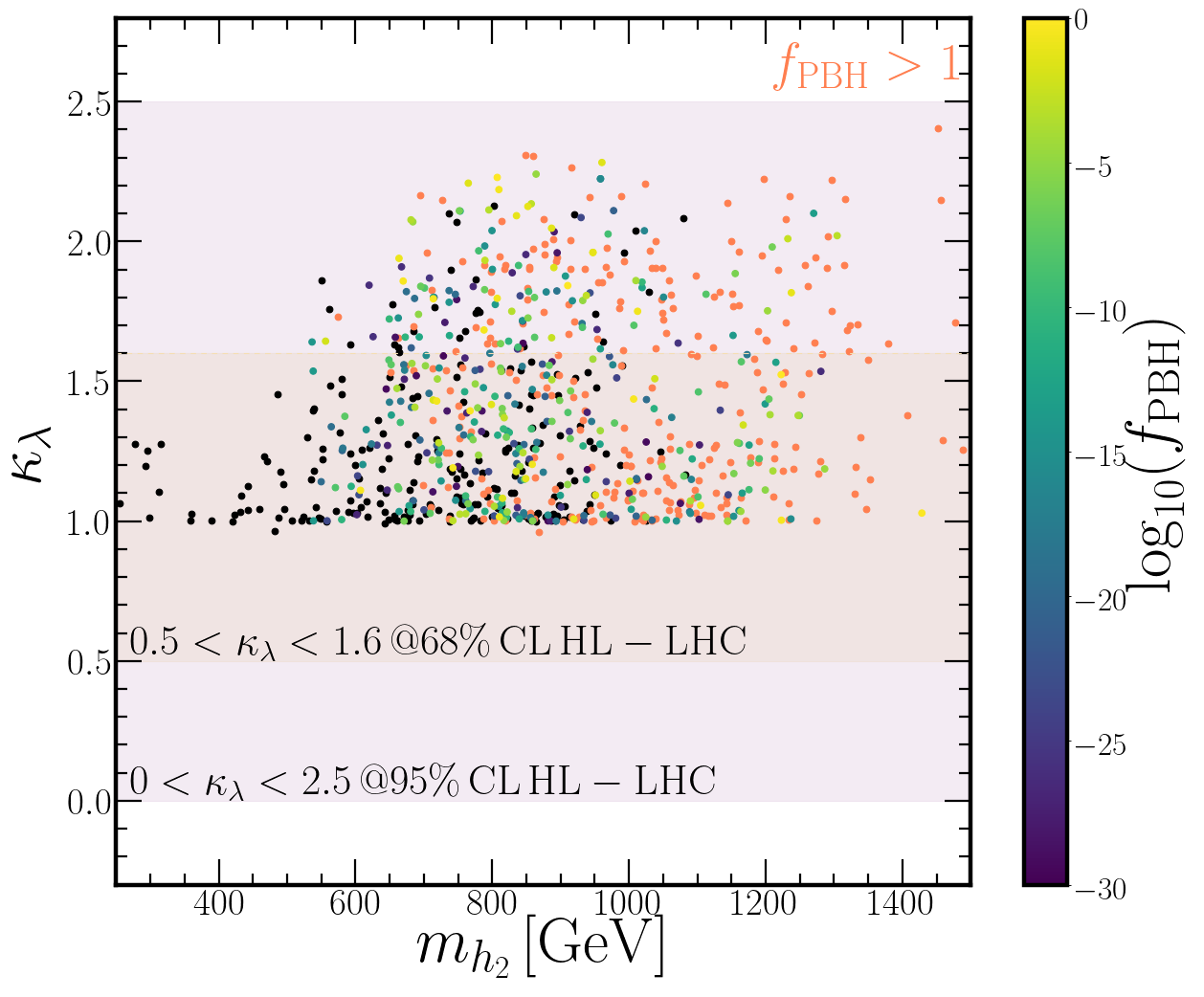}
    \includegraphics[height=0.38\textwidth]{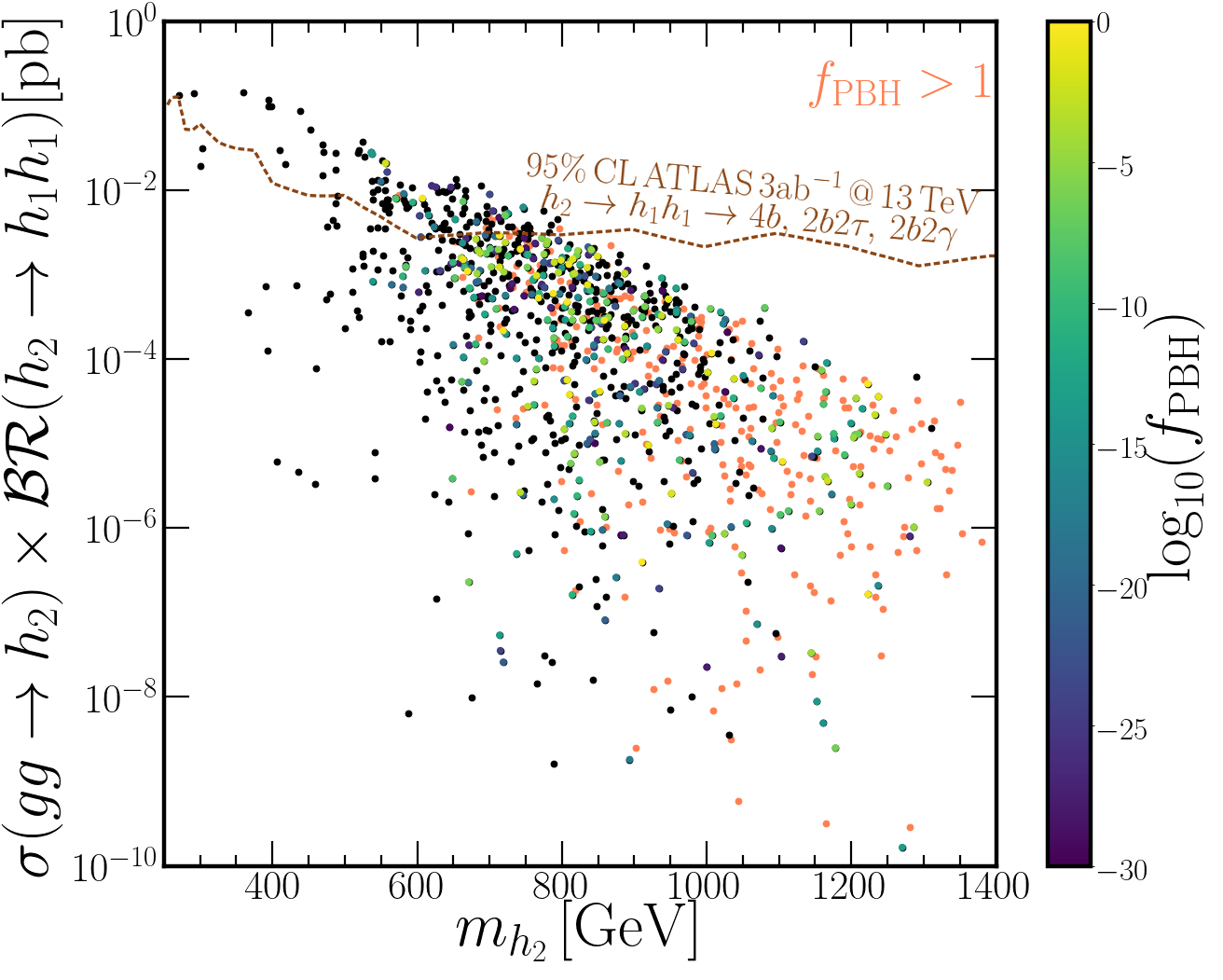}
    \caption{Left panel: The mass of the heavy scalar $m_{h_2}$ versus the triple Higgs coupling modifier $\kappa_\lambda=\lambda_{111}/\lambda_{111}^\mathrm{SM}$. The dashed curve represents the projected sensitivity at the HL-LHC with an integrated luminosity of 3~ab$^{-1}$, shown at the $1\sigma$ and $2\sigma$ levels~\cite{ATL-PHYS-PUB-2022-053}. Right panel: The HL-LHC projected limits for the cross-section of the heavy Higgs decaying into di-Higgs, $h_2 \rightarrow h_1 h_1$, are represented by the dashed curve~\cite{ATLAS:2023vdy}.  The points are color-coded by $f_{\mathrm{PBH}}$, and the orange color denotes the parameter excluded from DM overproduction $f_{\mathrm{PBH}}>1$. The black represent parameter points with $f_{\rm PBH}$ in the range between $10^{-300}$ and $10^{-30}$. As discussed in \autoref{sec:GW}, the formation of at least one PBH in the observable Universe imposes a lower bound of $f_\mathrm{PBH} > 10^{-30}$.}
    \label{fig:mH2_kappa}
\end{figure}

 PBH formation is favored in regions of parameter space leading to a large barrier at zero temperature, requiring specific cubic couplings, as discussed in~\autoref{sec:PBH_xSM}. This results in distinct triple Higgs couplings $\lambda_{111}$, which can provide relevant LHC phenomenology, as we will demonstrate.  In the left panel of~\autoref{fig:mH2_kappa}, we show triple Higgs coupling modifier, $\kappa_\lambda=\lambda_{111}/\lambda_{111}^\mathrm{SM}$, as a function of the heavy Higgs mass $m_{h_2}$. The projected sensitivity at the HL-LHC, with 3~ab$^{-1}$ integrated luminosity, ranges between $0.5 < \kappa_\lambda < 1.6$ at the $1\sigma$ level  and $0 < \kappa_\lambda < 2.5$ at the $2\sigma$ level.  This sensitivity is based on a combination of the $b\bar{b}b\bar{b}$, $b\bar{b}\gamma\gamma$, and $b\bar{b}\tau^+\tau^-$ channels~\cite{ATL-PHYS-PUB-2022-053}. Notably, for the PBH formation parameter space, $\kappa_\lambda$ tends to exceed the unity. This is evident when expanding the coupling $\lambda_{111}$ in~\autoref{eq:L111}, using a Taylor series for small values of $\theta$~\cite{Alves:2018jsw},
\begin{equation}
\lambda_{111}=\frac {m^2_{h_1}}{2v_\mathrm{EW}}\left [ 1+\theta^2\left ( -\frac {3}2+\frac {2m^2_{h_2}-2b_3v_s-4b_4v_s^2}{m^2_{h_1}} \right )+\mathcal{O}(\theta^3) \right ].
\end{equation}
The preference for negative values of the cubic coupling $b_3$ and a tendency towards $\Theta_{hs}^{-1}\sim-0.55$ ensure that  $\kappa_\lambda\gtrsim 1$. The HL-LHC will be sensitive, at the one to two $\sigma$ level, to a substantial fraction of the parameter points that allow for PBH formation, using the triple Higgs coupling constraints.

Resonant di-Higgs searches also provide relevant probes for parameter space regime that favors PBHs. In the right panel of~\autoref{fig:mH2_kappa}, we present the heavy Higgs cross section decaying to di-Higgs, $h_2 \rightarrow h_1 h_1$, as a function of heavy Higgs mass $m_{h_2}$. The ATLAS collaboration obtained the $95\%$ confidence level limit for this channel by combining the final states $4b$, $2b 2\tau$, and $2b 2\gamma$ in Ref.~\cite{ATLAS:2023vdy}. The dotted line represents this result scaled to the high-luminosity LHC with $\mathcal{L}=3~\mathrm{ab}^{-1}$ of data. The signal cross-section for Higgs production was obtained at NNLO+NNLL QCD and incorporates the effects of top and bottom quark masses up to NLO~\cite{LHCHiggsCrossSectionWorkingGroup:2016ypw, cerntwiki, Alves:2020bpi}. The resonant di-Higgs channel can probe some of the parameter space that displays PBH formation with $m_{h_2}<800\,\mathrm{GeV}$  with relatively low PBH fraction.

\begin{figure}[!t]
\centering
    \includegraphics[height=0.43\textwidth]{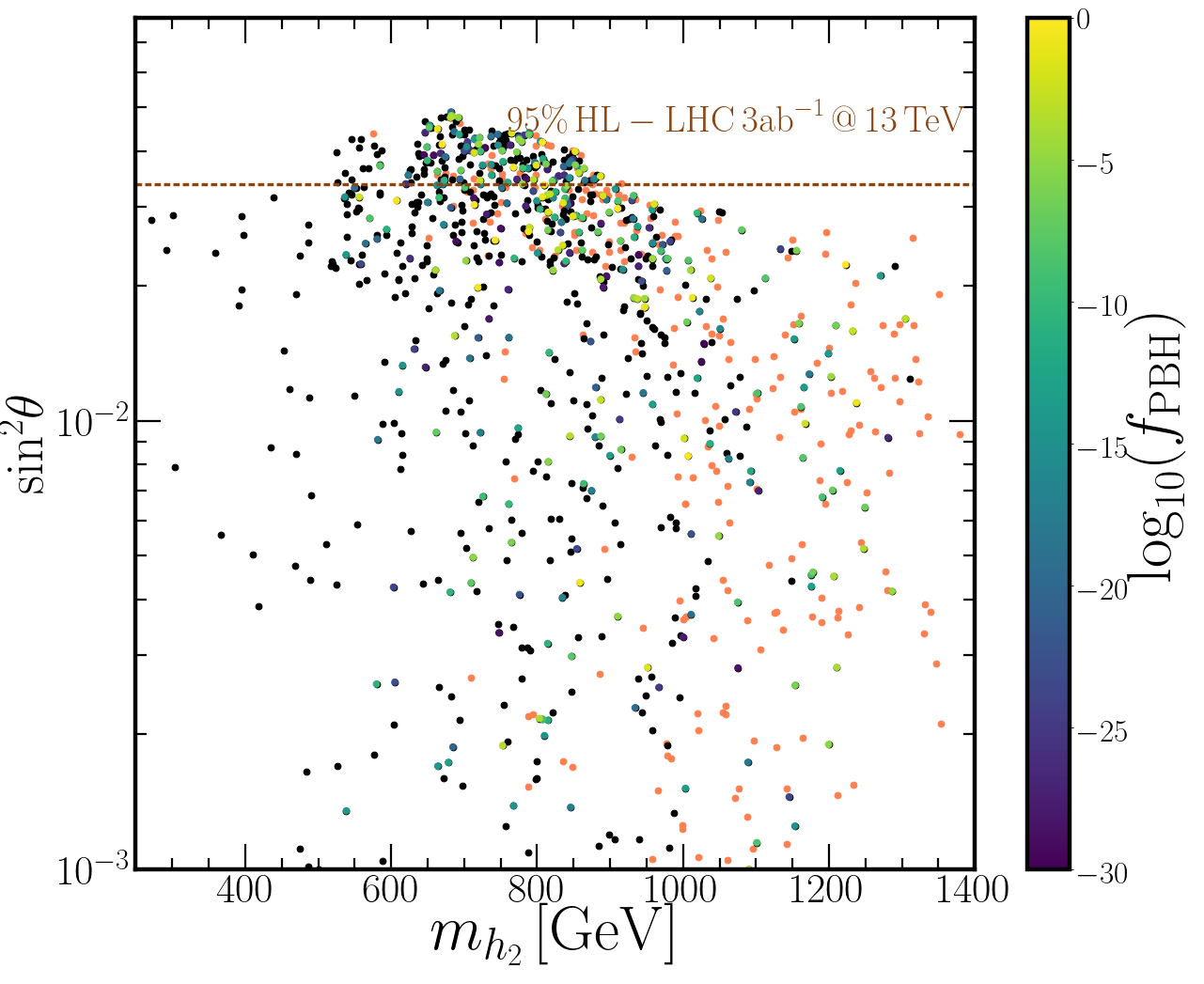}
    \caption{The mass of heavy scalar $m_{h_2}$ as a function of the sine-square of mixing angle $\sin^2 \theta$, with the HL-LHC projected sensitivity to the mixing angle indicated by the dashed line. The data points are color-coded according to $\text{log}_{10} (f_{\mathrm{PBH}})$, and the orange color represents the parameter excluded from DM overproduction, $f_{\mathrm{PBH}}>1$. Black points denote parameter points with $f_{\rm PBH}$ in the range between $10^{-300}$ and $10^{-30}$.}
    \label{fig:mH2_st2}
\end{figure}
Higgs signal strength measurements across various channels require the coupling to $h_1$  to be nearly identical to the SM Higgs coupling. In this model, the coupling of  $h_1$  to SM particles is simply the corresponding SM coupling scaled by  $\cos\theta$. At the HL-LHC, Higgs signal strength measurements are expected to exclude $\sin^2 \theta>0.034$~\cite{EuropeanStrategyforParticlePhysicsPreparatoryGroup:2019qin} at 95\% confidence level. In~\autoref{fig:mH2_st2}, we show the mass of heavy scalar $m_{h_2}$ as a function of the sine-square of mixing angle $\sin^2 \theta$, with the dashed line denoting the projected sensitivity at HL-LHC. The Higgs signal strength measurement can constrain PBH parameter space for $m_{h_2}$ in the range of $500-900\,\mathrm{GeV}$.

\begin{figure}[t!]
\includegraphics[width=0.48\textwidth]{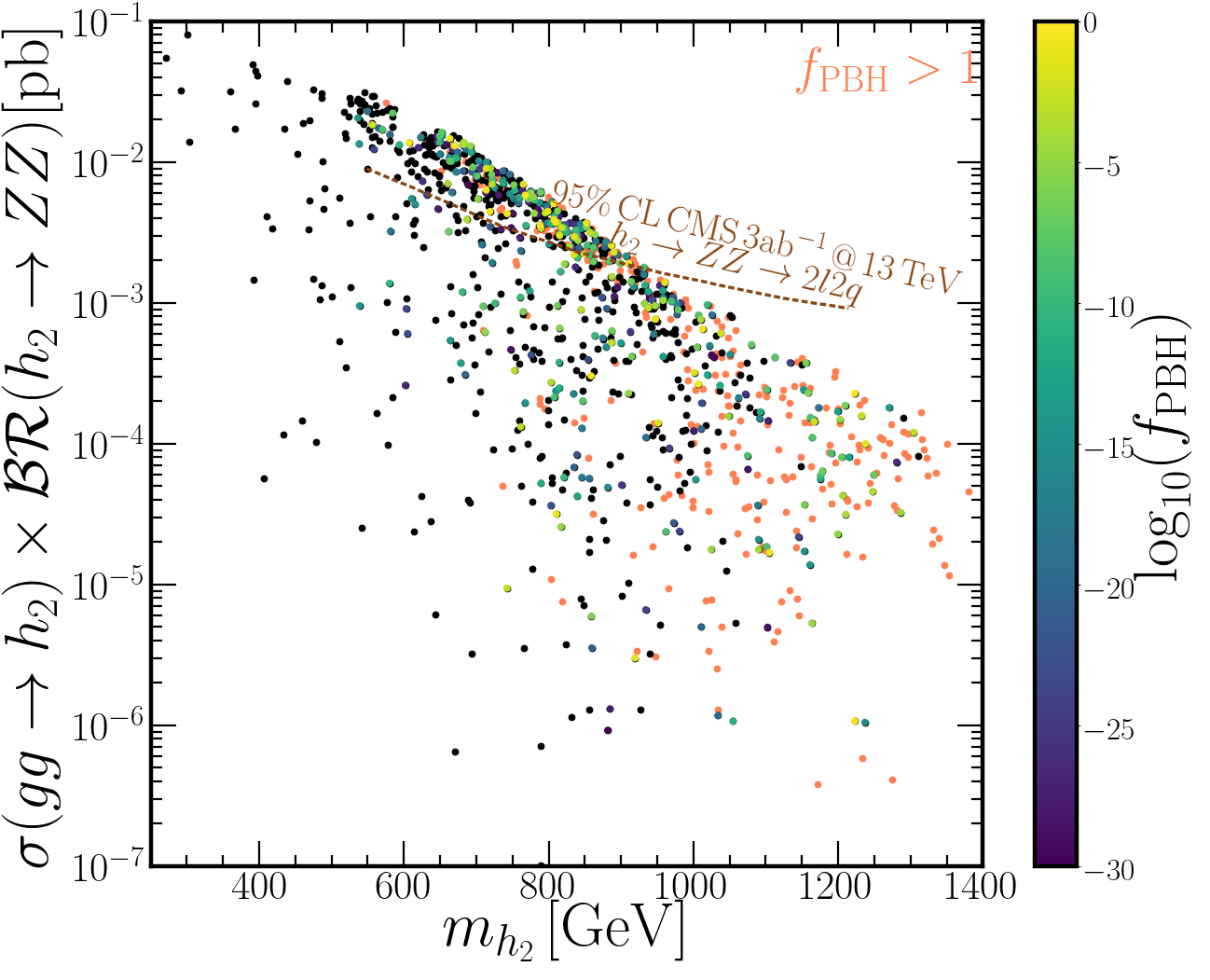}
\includegraphics[width=0.48\textwidth]{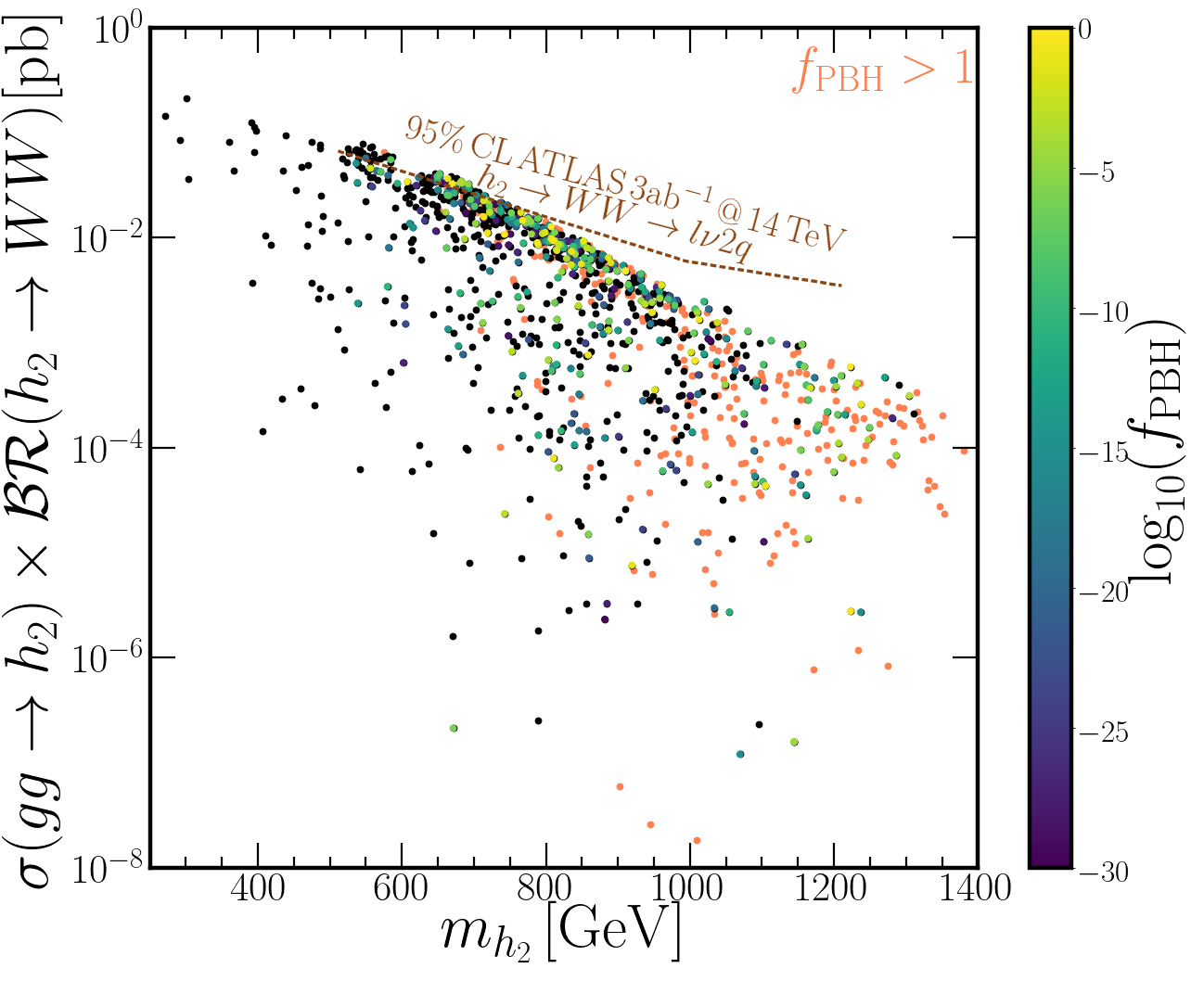}
\caption{The mass of the heavy scalar $m_{h_2}$ versus , di-boson $h_2 \rightarrow ZZ$ (left panel) and $h_2 \rightarrow WW$ (right panel) color coded by $f_{\mathrm{PBH}}$ and, the orange color denotes the parameter excluded due to overproduction of DM $f_{\mathrm{PBH}}>1$. The black denotes parameter points with $f_{\rm PBH}$ in the range between $10^{-300}$ and $10^{-30}$. The dashed lines in the figure represent the HL-LHC sensitivity projections from ATLAS and CMS collaborations~\cite{CMS:2019qzn, ATL-PHYS-PUB-2018-022}.}
\label{fig:gg_VV}
\end{figure}

The resonant di-boson $ZZ/WW$ searches can also lead to relevant constraints. In~\autoref{fig:gg_VV}, we present the sensitivity to the PBH parameter space from the di-boson channels $h_2 \rightarrow ZZ$ (left panel) and $h_2 \rightarrow WW$ (right panel). In Refs.~\cite{CMS:2019qzn, ATL-PHYS-PUB-2018-022}, the ATLAS and CMS collaborations analyzed the projected sensitivities of the HL-LHC to heavy Higgs resonant searches in the channels $pp \rightarrow h_2 \rightarrow ZZ \rightarrow 2\ell 2q$ at $\sqrt{s}=13\, \mathrm{TeV}$ and $pp \rightarrow h_2 \rightarrow WW \rightarrow \ell \nu 2q$ at $\sqrt{s}=14\, \mathrm{TeV}$. The results are depicted as dotted lines in~\autoref{fig:gg_VV}. While the $h_2 \rightarrow WW$ channel can probe PBH formation parameter space with $m_{h_2}\lesssim 700~\rm GeV$, the $h_2 \rightarrow ZZ$ channel offers significantly more sensitivity to PBH formation parameter space at HL-LHC with $m_{h_2}\lesssim 900~\rm GeV$.

\section{Conclusion}
\label{sec:conclusion}

In this paper, we considered the PBH formation during the electroweak phase transition in the singlet extension of the SM, xSM. PBH formation during phase transition requires sufficient supercooling such that the probability of having a {\it late patch}, where the system remains in the false vacuum, is large. In order to achieve that, it is found that the $S_3/T$ behavior as a function of temperature should have a U-shape, which provides a strict requirement on the evolution of the potential. By scrutinizing the evolution of the barrier height and depth of the true vacuum, we find a strong correlation between the shape of the effective potential with the PBH formation.

PBHs, which can serve as dark matter components~\cite{Green:2020jor}, can leave observable signals in microlensing experiments due to their gravitational effects. The mass of these PBHs is restricted to a narrow range, approximately $10^{-5}$ solar mass, which is about the mass of Earth. This constraint is influenced by the scale of the electroweak phase transition in the xSM. Furthermore, our analysis shows that this model can produce a broad spectrum of dark matter fractions. Notably, the current OGLE experiment has observed six ultrashort timescale events consistent with Earth-mass PBHs. Hence, PBH formation through supercooled electroweak phase transition provides a promising explanation.

Additionally, we find that the peak frequency of the gravitational wave induced by such a supercooled phase transition can be naturally probed by the future LISA mission with sufficient signal strength because of its supercooled nature. LISA's potential to probe the parameter space region associated with PBHs formation in the xSM across the observable Universe positions it to either confirm the proposed PBH scenario or definitively refute it.

The formation of PBHs impose stringent constraints on the parameter space of xSM model, particularly concerning the cubic terms that induce the barrier in the effective potential. Collider searches emerge as valuable tools for investigating these regions, capable of encompassing significant portions of the parameter space, even with low PBH fractions $f_{\rm PBH}$. Furthermore, in the event of PBH discovery, collider searches have the potential to scrutinize the underlying theoretical model. 

Thus, the study of PBHs from supercooled electroweak phase transitions offers a powerful and complementary interplay between different experimental approaches. Microlensing observations can detect the gravitational lensing signatures of PBHs, while gravitational wave experiments, such as LISA, can probe the phase transition dynamics responsible for PBH formation. Collider searches, in turn, offer a crucial window into the electroweak-scale physics underlying these phenomena. Together, these experimental approaches offer a comprehensive strategy for probing the PBHs generated by supercooled electroweak phase transitions.

\begin{acknowledgments}
We would like to thank Iason Baldes, Sumit Biswas, Bhaskar Dutta, Shinya Kanemura, Ian Lewis, Philip Lu, Mar\'ia Olalla Olea-Romacho, Masanori Tanaka, Ke-Pan Xie, and Tao Xu for helpful discussions. D.G. thanks the group at the IPPP-Durham University for hosting him during the final stages of this project. D.G. and A.K. are supported by the U.S. Department of Energy under grant number DE-SC 0016013. Y.W. is supported by the National Natural Science Foundation of China (NNSFC) under grant No.~12305112. Some computing for this project was performed at the High Performance Computing Center at Oklahoma State University, supported in part through the National Science Foundation grant OAC-1531128.
\end{acknowledgments}

\appendix
\section{Parametrization scan for xSM}
\label{app:paramscan}
This appendix illustrates how we scan the xSM scalar potential parameter space, satisfying current phenomenological constraints. The scalar potential in~\autoref{eq:V0} can be expressed differently, where EW vacuum has VEV $\left \langle S \right \rangle =0$ by translating the coordinate system~\cite{Lewis:2017dme, Liu:2021jyc}. In this basis, scalar potential has an additional contribution of tadpole term $b_1^\prime S$,
\begin{align}
V(H,S)=&-\mu^{\prime 2} H^\dagger H+\lambda^\prime (H^\dagger H)^2+\frac {a^\prime_1}{2}H^\dagger HS \nonumber\\
&+\frac {a^\prime_2}{2}H^\dagger H S^2+b_1^\prime S+\frac {b^\prime_2}{2}S^2+\frac{b^\prime_3}{3}S^3+\frac{b^\prime_4}{4}S^4,
\label{eq:V0app}
\end{align}
The minimization condition at EW vacuum $(v_{\mathrm{EW}},0)$ provides,
\begin{equation}
\mu^{\prime2}=\lambda^\prime v_{\mathrm{EW}}^2\, \quad \quad
b_1^\prime=-\frac{v_{\mathrm{EW}}^2}{4}a_1^\prime,
\end{equation}
and the remaining coefficients can be expressed in terms of the scalar masses and mixing angle,
\begin{equation}
a_1^\prime=\frac {s_{2\theta}}{v_{\mathrm{EW}}}(m_{h_1}^2-m_{h_2}^2), \quad
b_2^\prime+\frac {a_2^\prime}{2}v_{\mathrm{EW}}^2=m_{h_1}^2s_\theta^2+m_{h_2}^2c_\theta^2, \quad
\lambda^\prime=\frac {1}{2v_{\mathrm{EW}}^2}(m_{h_1}^2c_\theta^2+m_{h_2}^2s_\theta^2).
\end{equation}
Consequently, we can use the following parameters as input parameters
\begin{equation}
\left \{ m_{h_2},\theta,a_2^\prime,b_3^\prime,b_4^\prime \right \}.
\end{equation}
We conducted a random uniform scan over the parameter region,
\begin{align}
    m_{h_2}&\in (130,1500)\,{\rm GeV},                &\theta&\in (-0.35,0.35)\,,  \quad                     \frac {b_{3}^\prime}{v_{\mathrm{EW}}}\in(-4\pi, 4\pi ),
    \nonumber \\
     b_4^\prime &\in (0.001,\frac {4\pi}{3})\,, & a_2^\prime&\in(-2\sqrt{\lambda^\prime b_4^\prime},4\pi).
     \label{eq:param_scan}
\end{align}
To obtain scalar potential in the form of~\autoref{eq:V0}, we shift $s \rightarrow s+\sigma$ and the coefficients transforms as
\begin{align}
\mu^2&=\mu^{\prime 2}-\frac {1}{2}a_2^\prime \sigma^2-\frac {1}{2}a_1^\prime \sigma, \\
 a_1&=a_1^\prime +2a_2^\prime \sigma, \\
b_1&=b_1^\prime +b_4^\prime \sigma^3+b_3^\prime \sigma^2+b_2^\prime \sigma,\\
b_2&= b_2^\prime +3b_4^\prime \sigma^2+2b_3^\prime\sigma, \\
b_3&=b_3^\prime +3b_4^\prime \sigma.
\end{align}
The coefficients $a_2,\lambda,$  and  $b_4$ remain the same as their counterparts and we obtain $\sigma$ such that $b_1=0$. The shifted $s$ obtains a VEV $v_\mathrm{s}=-\sigma$. If $v_\mathrm{s}$ is negative, we can ensure that $v_\mathrm{s}$ remains positive by the $Z_2$ transformation $v_\mathrm{s} \rightarrow - v_\mathrm{s}$, $s \rightarrow - s$, $a_1 \rightarrow - a_1$ and  $b_3 \rightarrow - b_3$.{\footnote{{  At finite temperatures, a term proportional to $T^2s$ can arise from  tree-level operators. However, as demonstrated in Refs.~\cite{Profumo:2007wc,Profumo:2014opa}, this contribution is suppressed by the mixing angle $\sin \theta$ and is numerically
numerically subdominant compared to the leading tree-level terms. Hence, we neglect this term in our analysis.}}}

\section{Phenomenological constraints}
\label{app:constraints}

In this appendix, we present the theoretical and experimental constraints used in this analysis~\cite{Profumo:2014opa, Chen:2014ask, Robens:2015gla, Buttazzo:2015bka, Chalons:2016jeu, Chen:2017qcz}.
\begin{itemize}
    \item \textbf{Boundedness condition:} The scalar potential in~\autoref{eq:V0} must be bounded below. Demanding this for arbitrary field directions provides the condition~\cite{Lewis:2017dme}
    \begin{equation}
\lambda>0, \quad b_4>0, \quad a_2>-\sqrt{4\lambda b_4}.
    \end{equation}
    \item \textbf{EW vacuum stability:} To guarantee that the EW vacuum is stable at zero temperature, one should ensure that no deeper minimum exists in the potential. We determine the nature of each minimum by solving ${\partial V}/{\partial \phi_i}=0,\,\left \{ \phi_1=h,\,\phi_2=s \right \}$ and evaluating the eigenvalues of the Hessian matrix ${\partial^2 V}/{\partial \phi_i \phi_j}$.
    \item \textbf{Perturbative unitarity:} Perturbative unitarity constraints the behavior of particle scatterings at high energies. At tree level, it requires all partial wave amplitudes $a_\ell (s)$ for all $2\to 2$ scattering processes to satisfy $\mathrm{Re}\,a_\ell (s)\leq  \frac 1 2$ for $\sqrt{s}\rightarrow \infty$~\cite{PhysRevD.16.1519}. Here, we include all scalar/vector boson $2\to 2$ scattering channels at the leading order. The details of the evaluation of the S-matrix are provided in Appendix~\ref{app:unitarity}.
    \item \textbf{Higgs signal strength measurements:} The 125~GeV Higgs signal measurements in various production and decay channels require the couplings of $h_1$ to be close to the SM Higgs.
In this analysis, we use HiggsSignals~\cite{Bechtle:2020uwn, Bechtle:2013xfa}\footnote{We used HiggsTools~\cite{Bahl:2022igd} to perform the HiggsBounds and HiggsSignals analysis in the singlet framework.} to apply the Higgs coupling constraints from ATLAS and CMS.

    \item \textbf{Heavy Higgs searches:} The heavy SM-like Higgs boson searches in ATLAS and CMS can provide a probe for $h_2$~\cite{ATLAS:2016paq, CMS:2016jvt, CMS:2016cma, ATLAS:2015sxd, CMS:2015hra, CMS:2013vyt, ATLAS:2015oxt, ATLAS:2015pre}. The partial decay width of heavy Higgs scalar decay into SM particles is given by
\begin{equation}
\Gamma_{h_2\rightarrow \text{SM}}=s_{\theta}^2\Gamma^{\mathrm{SM}}(m_{h_2}),
\end{equation}
where $\Gamma^{\mathrm{SM}}(m_{h_2})$ is the SM Higgs decay width evaluated at $m_{h_2}<2m_{h_1}$~\cite{LHCHiggsCrossSectionWorkingGroup:2013rie}. In the regime $m_{h_2}>2m_{h_1}$, the decay mode $h_2 \rightarrow h_1 h_1$ is kinematically allowed. The corresponding partial decay width is given by~\cite{Huang:2017jws, Robens:2015gla}
\begin{equation}
\Gamma_{h_2\rightarrow h_1h_1}=\frac {\lambda_{211}^2\sqrt{1-\frac{4m_{h_1}^2}{m_{h_2}^2}}}{8\pi m_{h_2}}.
\end{equation}
We used HiggsBounds~\cite{Bechtle:2008jh, Bechtle:2011sb, Bechtle:2013wla, Bechtle:2020pkv} to implement the current experimental constraints.
\item \textbf{Electroweak precision measurements:} The $W$ boson mass measurement~\cite{Lopez-Val:2014jva} and oblique EW corrections~\cite{PhysRevD.46.381, Hagiwara:1994pw} put additional constraints on the parameter space of the xSM. The $W$ boson mass $m_W$ depends on the loop corrections of the vector boson self-energies~\cite{Lopez-Val:2014jva}. In the xSM model, the reduced Higgs coupling and the presence of heavier scalar can modify the loop corrections, and these modifications only depend on two parameters $\theta$ and $m_{h_2}$. The same parameters enter the corrections to the oblique parameters $S,\,T,$ and $U$. The $W$ mass experimental value $m_W^{\mathrm{exp}}= 80.3545 \pm 0.0059\,\mathrm{GeV}$~\cite{10.1093/ptep/ptac097}
puts a more stringent bound on these parameters than the oblique corrections~\cite{Lopez-Val:2014jva,Robens:2015gla}.
\end{itemize}

\section{Perturbative unitarity of the S-matrix}
\label{app:unitarity}
For scalars (and longitudinal gauge boson) $2\to 2$ process, the scattering amplitude can be written as
\begin{equation}
A(\alpha)=16 \pi \sum_{J}(2J+1)a_JP_J(\cos \alpha),
\end{equation}
where $\alpha$ denotes scattering angle and $P_J$ is the Legendre polynomial. The perturbative unitarity demands $|Re(a_{0}^{max})| \leq \frac 12$~\cite{Lee:1977eg}. There are seven neutral ($h_1h_1,\, h_2h_2,\, h_1h_2,\, h_1Z,\, h_2Z,\,ZZ,\,W^+W^-$), three singly-charged ($h_1W^+,\, h_2W^+,\,ZW^+$) and one doubly-charged ($W^+ W^+$) channels. Therefore, the scattering amplitude matrix $a_0$ can be written as a direct sum of matrices $a_0=\mathcal{S}_1\mathop{{\bigoplus}} \mathcal{S}_2 \mathop{{\bigoplus}} \mathcal{S}_3$ from these three groups~\cite{Alves:2018jsw}. The perturbative unitarity condition translates into the magnitude of the largest eigenvalue should be less than $8\pi$.

The non-zero elements of the neutral $7 \times 7$ submatrix $\mathcal{S}_1$ are given by~\cite{Alves:2018jsw,Kanemura:2015ska}
\begin{align}
\mathcal{S}_{11}&=-3(a_2c_\theta^2s_\theta^2+b_4s_\theta^4+\lambda c_\theta^4),\\
\mathcal{S}_{12}&=\frac {1}{8}(3c_{4\theta}(-a_2+b_4+\lambda)-a_2-3b_4-3\lambda),\\
\mathcal{S}_{13}&=\frac {3s_{2\theta}(c_{2\theta}(-a_2+b_4+\lambda)-b_4+\lambda)}{2\sqrt{2}},\\
\mathcal{S}_{16}&=-\frac 12 a_2 s_\theta^2-\lambda c_\theta^2,\\
\mathcal{S}_{17}&=-\frac {a_2 s_\theta^2+2\lambda c_\theta^2}{\sqrt{2}},\\
\mathcal{S}_{22}&=-3(a_2 c_\theta^2s_\theta^2+b_4 c_\theta^4+\lambda s_\theta^4),\\
\mathcal{S}_{23}&=-\frac {3s_{2\theta}(c_{2\theta}(-a_2+b_4+\lambda)+b_4-\lambda)}{2\sqrt{2}},\\
\mathcal{S}_{26}&=-\frac {1}{2}a_2 c_\theta^2-\lambda s_\theta^2,\\
\mathcal{S}_{27}&=-\frac {a_2 c_\theta^2+2\lambda s_\theta^2}{\sqrt{2}},\\
\mathcal{S}_{33}&=\frac {1}{4}(3c_{4\theta}(-a_2+b_4+\lambda)-a_2-3b_4-3\lambda),\\
\mathcal{S}_{36}&=\frac {(2\lambda-a_2)c_\theta s_\theta}{\sqrt{2}},\\
\mathcal{S}_{37}&=(2\lambda-a_2)c_\theta s_\theta,\\
\mathcal{S}_{44}&=-a_2s_\theta^2-2\lambda c_\theta^2,\\
\mathcal{S}_{45}&=(2\lambda-a_2)c_\theta s_\theta,\\
\mathcal{S}_{55}&=-a_2c_\theta^2-2\lambda s_\theta^2,\\
\mathcal{S}_{66}&=-3\lambda,\\
\mathcal{S}_{67}&=-\sqrt{2}\lambda,\\
\mathcal{S}_{77}&=-4\lambda.
\end{align}
For singly-charged channel the submatrix $\mathcal{S}_2$ is given by
\begin{equation}
    \mathcal{S}_2=\begin{bmatrix}
-2\lambda c_\theta^2-a_2s_\theta^2 & (2\lambda-2)c_\theta s_\theta & 0\\
(2\lambda-2)c_\theta s_\theta & -a_2 c_\theta^2-2\lambda s_\theta^2 & 0\\
0 & 0 & -2\lambda
\end{bmatrix}.
\end{equation}
We have only one process for the doubly-charged channel and the submatrix is $\mathcal{S}_3=-2\lambda$.

\section{PBH calculation}
\label{app:PBH_calc}
In the following discussion, it is implied that quantities without specific indices are defined within the context of the false vacuum. Using $dt=-\frac{dT}{HT}$, we can express~\autoref{eq:Fin} in terms of temperature as~\cite{Ellis:2018mja,Baldes:2023rqv}
\begin{equation}
F_\mathrm{in}(T)=\mathrm{exp}\left [ -\frac {4\pi}{3}\int _{T}^{T_i}\frac {dT^\prime \Gamma(T^\prime)}{T^{\prime 4}H(T^\prime)^3}\left ( \int_T^{T^\prime} \frac {A_\mathrm{in}(T^\prime,\tilde{T})d\tilde{T}}{H(\tilde T)}\right )^3 \right ],
\label{eq:Fin(t)}
\end{equation}
where $H$ is the Hubble rate at false vacuum $H_{\mathrm{false}}$ and we define $T$ as the temperature of the false vacuum. The Hubble rate at the false vacuum is obtained from the Friedmann equation including the vacuum and radiation energy inside the false vacuum:
\begin{equation}
    H(T)^2\equiv H_{\mathrm{false}}(T)^2=\frac {1}{3M_\mathrm{Pl}^2}\left (\Lambda_{\mathrm{vac}}(T)+ \frac{\pi^2}{30}g_\star T^4  \right ).
\end{equation}
Furthermore, the function $A_\mathrm{in}(T^\prime,\tilde{T})$ in \autoref{eq:Fin(t)}is defined as
\begin{equation}
A_\mathrm{in}(T^\prime, \tilde{T})=\frac {a_\mathrm{in}(T^\prime)T^\prime}{a_\mathrm{in}(\tilde{T})\tilde{T}}.
\label{eq:Ainttp}
\end{equation}
The Friedman equation governing the evolution of  Hubble rate,  radiation energy density, vacuum energy density, and scale factor of the {\it late patch} can be rewritten as a function of the temperature as
\begin{subequations}
\begin{align}
H_{\mathrm{in}}^2&=\frac {1}{3M^2_{\mathrm{Pl}}}\left ( \rho_R^{\mathrm{in}}+\rho_V^{\mathrm{in}} \right ),\\
\frac {d\rho_R^{\mathrm{in}}}{dT}&=\frac{4H_{\mathrm{in}}\rho_R^{\mathrm{in}}}{HT}- \frac {d\rho_V^{\mathrm{in}}}{dT},\\
\frac {da_{\mathrm{in}}}{dT}&=-\frac{a_{\mathrm{in}}}{T}\frac{H_{\mathrm{in}}}{H},
\end{align}
\label{eq:inevol}
\end{subequations}
where vacuum energy density is given by $ \rho^{\mathrm{in}}_{V}=F_{\mathrm{in}}(T) \Lambda_{\mathrm{vac}}(T)$.
The temperature evolution of $\rho^{\mathrm{in}}$, the scalar factor $a_{\mathrm{in}}$ for this region can be solved numerically using~\autoref{eq:inevol} with initial conditions of the false vacuum which are given by
\begin{equation}
\rho^{\mathrm{in}}_{V}(T_i)=\Lambda_{\mathrm{vac}}(T_i),\quad \rho^{\mathrm{in}}_{R}(T_i)= \frac{\pi^2}{30}g_\star T_i^4, \quad a_{\mathrm{in}}(T_i)=\frac{1}{T_i}.
\end{equation}
In a similar approach, we can consider the evolution of the {\it background region}, where we can assume for the surrounding region that the first bubble nucleates around critical temperature $T_c$,
and the average spatial fraction of the false vacuum in such {\it background region} is given by~\footnote{The critical temperature $T_c$ is defined as the point at which the false and would-be true vacua degenerate.}
\begin{equation}
    F_\mathrm{out}(T)=\mathrm{exp}\left [ -\frac {4\pi}{3}\int _{T}^{T_c}\frac {dT^\prime \Gamma(T^\prime)}{T^{\prime 4}H(T^\prime)^3}\left ( \int_T^{T^\prime} \frac {A_\mathrm{out}(T^\prime,\tilde{T})d\tilde{T}}{H(\tilde T)}\right )^3 \right ],
\end{equation}
where the function $A_{\mathrm{out}}(T^\prime, \tilde{T})$ is defined similar to~\autoref{eq:Ainttp} by replacing $a_{\mathrm{in}}$ with $a_{\mathrm{out}}$ and all quantities without specific indices are defined at false vacuum. The evolution of Hubble rate, scale factor, radiation energy density and vacuum energy density of the {\it background region} can be solved numerically using a similar equation of~\autoref{eq:inevol}, by substituting indices from ``in" to ``out" and with the initial condition given by
\begin{equation}
\rho^{\mathrm{out}}_{V}(T_c)=\Lambda_{\mathrm{vac}}(T_c),\quad \rho^{\mathrm{out}}_{\mathrm{rad}}(T_c)=\frac{\pi^2}{30}g_\star T_c^4, \quad a_{\mathrm{out}}(T_c)=\frac {1}{T_c}.
\end{equation}
The probability that no bubble nucleates in the past wall cone of the Hubble volume at $T>T_i$ is given by~\cite{Kawana:2022olo, Liu:2021svg,Baldes:2023rqv},
\begin{equation}
P(T_i)=\mathrm{exp}\left [ -\int_{T_i}^{T_c}\frac {dT^\prime \Gamma(T^\prime)}{T^\prime H(T^\prime)} a_{\mathrm{in}}(T^\prime)^3V_{\mathrm{coll}} \right ],
\label{eq:P_pbh_app}
\end{equation}
where the volume factor $V_{\mathrm{coll}}$ is given by
\begin{equation}
V_{\mathrm{coll}}=\frac {4\pi}{3}\left [ \frac {1}{a_{\mathrm{in}}(T_{\mathrm{PBH}})H_{\mathrm{in}}(T_{\mathrm{PBH}})}+\int_{T_\mathrm{PBH}}^{T^\prime}\frac {d\tilde{T}}{\tilde{T}H(\tilde{T})a_{\mathrm{out}}(\tilde{T})} \right ]^3.
\end{equation}

\section{Gravitational wave signatures}
\label{app:GW}
First-order phase transition (FOPT) occurring in the early Universe may lead to detectable stochastic gravitational wave (GW) signals in the present era~\cite{Caprini:2015zlo, Cai:2017cbj,Caprini:2018mtu}. Supercooling is anticipated for the parameter space, leading to PBH formation. Therefore, we designate the percolation temperature $T_p$ as the phase transition temperature, which serves as the point for evaluating gravitational waves. During FOPT,  GWs emerge from three distinct sources: bubble collisions, sound waves, and magnetohydrodynamic (MHD) turbulence. The GW spectrum, denoted as $\Omega_{\mathrm{GW}}(f)$, represents the current GW energy density per logarithmic frequency interval relative to the critical energy density of the Universe. Thus, $\int \Omega_{\mathrm{GW}}d\ln f$ signifies the fraction of GW energy density compared to the critical energy density of the Universe. The GW energy density can be linearly approximated as
\begin{equation}   \Omega_{\mathrm{GW}}h^2=\Omega_{\mathrm{coll}}h^2+\Omega_{\mathrm{SW}}h^2+\Omega _{\mathrm{turb}}h^2\,.
\label{eq:GW}
\end{equation}
Here, $h=H_0/100~\text{Km}/ \text{s}/\text{Mpc}$ represents the dimensionless Hubble parameter at the present time. Next, we will examine each of the components in~\autoref{eq:GW}.
\subsection{Sound waves}
The energy released during the phase transition to the plasma can be distributed between heat and fluid motion. Numerical estimations suggest that the energy-momentum tensor of the fluid following bubble collision bears resemblance to that of an ensemble of sound waves~\cite{Hindmarsh:2013xza, Hindmarsh:2017gnf, Hindmarsh:2016lnk}. Notably, these sound waves emerge as a significant source of gravitational waves. The redshifted peak amplitude of the gravitational waves stemming from these sound waves are given by~\cite{Hindmarsh:2017gnf, Athron:2023mer},
\begin{equation}
h^2\Omega_{\mathrm{sw}}=3\mathcal{R}_{\Omega}\left ( \frac{k_s \alpha}{\alpha+1} \right )^2\left ( \frac {H_\star R_\star}{c_{s,f}} \right )\frac {M(s,r_b,b)}{\mu_f(r_b)}\Upsilon(\tau_{\mathrm{sw}})\tilde{\Omega}_{\mathrm{GW}},
\label{eq:gw_sw}
\end{equation}
where the redshift factor for the amplitude $\mathcal{R}_{\Omega}$ is provided by,~\cite{Athron:2023mer}
\begin{equation}
R_{\Omega}=1.384\times 10^{33}\left ( \frac{\mathrm{GeV}^3}{s_1} \right )^{\frac {4}{3}}\left (\frac {H_\star} {\mathrm{Gev}}  \right )^2,
\end{equation}
where $s_1$ is the entropy at reheat temperature $T_{\mathrm{rh}}$ and $H_{\star}$ is the Hubble factor at gravitational wave temperature $T_p$. The reheating temperature can be approximated as,
\begin{equation}
T_{\mathrm{rh}}=(1+\alpha)^{\frac 1 4}T_p.
\end{equation}
To estimate mean bubble separation $R_\star$, we calculate the mean bubble number density considering that true vacuum bubbles can only nucleate within regions that remain in the false vacuum state. The mean bubble separation $R_{\star}$ is given by~\cite{PhysRevLett.44.963.2, Athron:2023xlk},
\begin{equation}
R_{\star}(T)\equiv \left (n_{B}(T)  \right )^{-\frac 13}=\left (\int_{T}^{T_c}dT^\prime\frac {\Gamma(T^\prime)F_{\mathrm{out}}(T^\prime)}{T^\prime H(T^\prime)}\left (\frac{a_{\mathrm{out}}(T^\prime)}{a_{\mathrm{out}}(T)}  \right )^3  \right )^{-\frac 13}.
\label{eq:Rstar}
\end{equation}

The spectral shape $M(s,r_b,b)$ is provided by~\cite{Hindmarsh:2017gnf, Athron:2023mer},
\begin{equation}
M(s,r_b,b)=s^9\left ( \frac {1+r_b^4}{r_b^4+s^4} \right )^{\frac {9-b}{b}}\left (\frac {b+4}{b+4-m+m^2}  \right )^{\frac{b+4}{2}},
\end{equation}
where,
\begin{align}
m=&\left ( 9r_b^4+b \right )/(r_b^4+1),\\
s=&f/f_p,\\
r_b=&f_b/f_p,\\
\mu_f(r_b)=&4.78-6.27r_b+3.34r_b^2.
\end{align}
In accordance with Table IV of~\cite{Hindmarsh:2017gnf}, we use $b=1$ and $\tilde{\Omega}_{\mathrm{GW}}=0.01$.  The $f_b$ and $f_p$ denote the redshifted frequencies corresponding to mean bubble separation and fluid shell thickness, which breaks power laws and are given by,
\begin{align}
f_b=&1.58\mathcal{R}_f\left ( \frac {1}{R_\star} \right )\left ( \frac {z_p}{10} \right )\\
f_p=&1.58\mathcal{R}_f\left ( \frac {1}{R_\star \Delta_w} \right )\left ( \frac {z_p}{10} \right ),
\end{align}
where $\Delta_w\approx \left | v_w-c_{s,f} \right |/v_w$ captures fluid shell thickness and $\mathcal{R}_f$ incorporates the redshift factor of the frequency,
\begin{equation}
\mathcal{R}_f=4.280\times 10^{11}\frac {\mathrm{Hz}}{\mathrm{Gev}}\left ( \frac {1\mathrm{GeV}^3}{s_1} \right )^\frac {1}{3}.
\end{equation}
We approximate speed of sound at false vacuum $c_{s,f}\approx 1/\sqrt{3}$~\cite{Athron:2023xlk} and $v_w$ is the Chapman-Jougat velocity~\cite{PhysRevD.25.2074,Giese:2020rtr},
\begin{equation}
v_w=\frac {1+\sqrt{3 \alpha(1+c^2_{s,f}(3\alpha-1))}}{c^{-1}_{s,f}+3\alpha c_{s,f}}.
\label{eq:vw}
\end{equation}
We consider the dimensionless wave number at the peak to be $z_p=10$, a value applicable to the supersonic detonations we are considering~\cite{Hindmarsh:2019phv, Hindmarsh:2017gnf}. The efficiency factor for the sound wave spectrum $k_s$ is provided by ~\cite{Caprini:2015zlo}
\begin{equation}
    k_s=\frac {\alpha}{0.73+0.083\sqrt{\alpha}+\alpha}.
\end{equation}
Recent studies ~\cite{Guo:2020grp, Hindmarsh:2020hop} demonstrate a suppression factor $\Upsilon (\tau_{\mathrm{SW}})$ arising from the finite active period $\tau_{\mathrm{SW}}$ of sound waves
\begin{equation}
    \Upsilon (\tau_{\mathrm{SW}})=1-\frac {1}{\sqrt{1+2\tau_{\mathrm{SW}} H_{\star}}}\,,
\end{equation}
where duration $\tau_{\mathrm{SW}}=R_\star/\bar{U}_f$ and represents the mean square velocity~\cite{Bodeker:2017cim}
\begin{equation}
    {\overline{U_f}}^2=\frac 34 \frac {\alpha}{1+\alpha}k_s.
\end{equation}
\subsection{Bubble collisions}
Nucleated bubbles, upon collision, lose their spherical symmetry, leading to the production of gravitational waves~\cite{Kosowsky:1991ua,Kosowsky:1992vn}. The redshifted peak amplitude due to bubble collisions is given by~\cite{Lewicki:2022pdb,Athron:2023mer},
\begin{equation}
\Omega_{\mathrm{coll}}=\mathcal{R}_{\Omega}A\left ( \frac {H_\star R_\star}{(8\pi)^{\frac 13}v_w} \right )^2\left ( \frac {k_c \alpha}{1+\alpha} \right )^2S_{\mathrm{coll}}(f),
\end{equation}
where the spectral shape is provided by,
\begin{equation}
S_{\mathrm{coll}}=\frac {(a+b)^c}{\left [ b\left ( \frac {f}{f_{\mathrm{coll}}} \right )^{-\frac ac}+a\left ( \frac {f}{f_{\mathrm{coll}}} \right )^{\frac bc} \right ]^c}.
\end{equation}
For the scenario where $T_{\tau \tau} \propto R^{-3}$, the fitting parameters $A$, $a$, $b$, and $c$ are available in Table I of Ref.~\cite{Lewicki:2022pdb}. The redshifted peak frequency is given by,
\begin{equation}
f_{\mathrm{coll}}=\mathcal{R}_f\left ( \frac {0.77(8\pi)^\frac{1}{3}v_w}{2\pi R_\star} \right )
\end{equation}

\subsection{MHD turbulence}

The energy injected into the plasma can trigger turbulence in the fluid, particularly in scenarios where the early Universe plasma exhibits an extremely high Reynolds number~\cite{Kamionkowski:1993fg}. This turbulent motion can be a significant source of gravitational waves~\cite{Witten:1984rs}. Moreover, turbulent motion within a fully ionized plasma can generate a turbulent magnetic field, further contributing to the production of gravitational waves. The redshifted peak amplitude stemming from turbulence can be parametrized by~\cite{Caprini:2010xv, Caprini:2009yp}
\begin{equation}
\Omega_{\mathrm{turb}}(f)=9.0\mathcal{R}_\Omega H_\star R_\star \left ( \frac {k_{\mathrm{turb}} \alpha}{1+\alpha} \right )^2S_{\mathrm{turb}}(f),
\end{equation}
where based on numerical simulation we adopt $k_{\mathrm{turb}}=0.05k_s$~\cite{Caprini:2015zlo} and the unnormalized spectral shape is given by
\begin{equation}
S_{\mathrm{turb}}(f)=\frac {(f/f_{\mathrm{turb}})^3}{\left ( 1+f/f_{\mathrm{turb}} \right )^\frac{11}{3}\left ( 1+8\pi f/\left ( \mathcal{R}_\star H_\star \right ) \right )},
\end{equation}
where $f_{\mathrm{turb}}=\mathcal{R}_f\frac {3.5}{R_\star}$ is the redshifted peak frequency~\cite{Caprini:2009yp}.

\bibliographystyle{refs}
\bibliography{references}

\end{document}